\documentclass[12pt]{article}
\usepackage[utf8]{inputenc}
\usepackage[english]{babel}
\usepackage{graphicx,amsfonts,amsmath,amssymb,amsthm,mathrsfs,mathabx}
\usepackage[usenames,dvipsnames]{xcolor}
\usepackage[toc,page]{appendix}
\usepackage{authblk}
\usepackage[bookmarksnumbered=true]{hyperref}
\usepackage[noend]{algpseudocode}

\usepackage[capitalise]{cleveref}
\crefname{equation}{}{}

\title{\vspace{-1cm} Bosonized Momentum Distribution of a Fermi Gas via Friedrichs Diagrams}

\author[1,*]{Sascha Lill}
\affil[1]{ORCID: \href{https://orcid.org/0000-0002-9474-9914}{0000-0002-9474-9914}, e--mail: \href{mailto:sascha.lill@unimi.it}{sascha.lill@unimi.it}}
\affil[*]{Universit\`a degli Studi di Milano, Dipartimento di Matematica, Via Cesare Saldini 50, 20133 Milano, Italy}

\usepackage[utf8]{inputenc}
\graphicspath{ {figures/}}
\usepackage{tikz}
\usetikzlibrary{decorations.pathreplacing, patterns}

\usepackage[section]{placeins}
\usepackage{capt-of, caption}
\usepackage{paralist}
\usepackage{booktabs}
\usepackage{hhline}
\usepackage{color}		

\newcommand{\bH}{\boldsymbol{H}}
\newcommand{\bK}{\boldsymbol{K}}
\newcommand{\bP}{\boldsymbol{P}}
\newcommand{\bQ}{\boldsymbol{Q}}
\newcommand{\balpha}{\boldsymbol{\alpha}}

\newcommand{\cC}{\mathcal{C}}

\newcommand{\cE}{\mathcal{E}}
\newcommand{\cF}{\mathcal{F}}

\newcommand{\cI}{\mathcal{I}}
\newcommand{\cJ}{\mathcal{J}}

\newcommand{\cU}{\mathcal{U}}

\newcommand{\CCC}{\mathbb{C}}

\newcommand{\NNN}{\mathbb{N}}
\newcommand{\RRR}{\mathbb{R}}

\newcommand{\TTT}{\mathbb{T}}
\newcommand{\ZZZ}{\mathbb{Z}}

\renewcommand{\a}{\textnormal{a}}
\newcommand{\ad}{\mathrm{ad}}
\newcommand{\bos}{\textnormal{b}}

\newcommand{\diam}{{\rm diam}}

\newcommand{\F}{{\rm F}}

\newcommand{\imag}{\mathrm{i}}
\newcommand{\nor}{\mathrm{nor}}

\newcommand{\sgn}{\mathrm{sgn}}

\newcommand{\supp}{\mathrm{supp}}

\newcommand{\kF}{k_\F}
\newcommand{\BF}{B_\F}

\newcommand{\tagg}[1]{ \stepcounter{equation} \tag{\theequation} \label{#1} } 

\usepackage{tikz}
\usetikzlibrary{decorations.pathreplacing, patterns}
\newcommand{\cont}[0]{
\!\! \raisebox{1pt}{
\begin{tikzpicture}
\filldraw[fill = white] (0, 0) circle (0.08);
\draw (-0.3,0) -- (-0.08,0);
\draw (0.3,0) -- (0.08,0);
\end{tikzpicture}
} \!\!
}
\newcommand{\contfer}[0]{
\!\! \raisebox{1pt}{
\begin{tikzpicture}
\filldraw[fill = white] (0, 0) circle (0.08);
\draw (-0.3,0) -- (-0.08,0);
\draw (0.3,0) -- (0.08,0);
\draw[thick] (0.3,-0.06) -- (0.15,-0.06);
\end{tikzpicture}
} \!\!
}

\newtheorem{theorem}{Theorem}

\newtheorem{proposition}{Proposition}

\theoremstyle{definition}
\newtheorem{remark}{Remark}
\theoremstyle{definition}

\theoremstyle{definition}

\theoremstyle{definition}

\newcommand\blfootnote[1]{%
  \begingroup
  \renewcommand\thefootnote{}\footnote{#1}%
  \addtocounter{footnote}{-1}%
  \endgroup
}

\begin{document}
\maketitle

\begin{abstract}
Recently \cite{benedikter2023momentum}, Benedikter and the author proved an approximate formula for the momentum distribution of a 3d fermionic gas interacting by a short-range pair potential in the mean-field regime, within a trial state close to the ground state. Here, we derive an exact formula for the momentum distribution in this trial state, using a diagrammatic formalism due to Friedrichs. We further demonstrate how the formula of Benedikter and the author arises from a restriction of the contributing diagrams to those corresponding to a bosonization approximation.
\end{abstract}



\tableofcontents

\section{Introduction}
\label{sec:intro}

In recent years, there has been a surge of interest in the mathematical research on three-dimensional (3d) fermionic quantum gases, employing bosonization techniques. Generally, such a system comprises $ N $ fermions on a 3d torus, $ \TTT^3 := [0, L]^3 $, whose state is described by a vector
\begin{equation}
\begin{aligned}
	&\qquad \psi \in L_{\a}^2(\TTT^{3N}) :=\\
	&\big\{ \psi \in L^2(\TTT^{3N}) \; \big\vert \; \psi( \ldots, x_i, \ldots, x_j, \ldots) = - \psi( \ldots, x_j, \ldots, x_i, \ldots) \; \forall 1 \le i < j \le n \big\} \;.
\end{aligned}
\end{equation}
The system is modeled by a Hamiltonian operator of the form
\begin{equation}
\label{eq:HN}
	H_N := \sum_{j = 1}^N - \hbar^2 \Delta_{x_j} + \lambda \sum_{i < j}^N V(x_i - x_j) \;,
\end{equation}
with a sufficiently regular pair potential $ V: \RRR^3 \to \RRR $, and where $ \hbar > 0 $ is the semiclassical constant and $ \lambda > 0 $ the coupling constant. For a given density $ \rho := \frac{N}{L^3} $, we require the energy to be extensive, which fixes $ \hbar \sim \rho^{-\frac 13} $ and $ \lambda \sim \rho^{-1} $.\\

\vspace{-0.5cm}
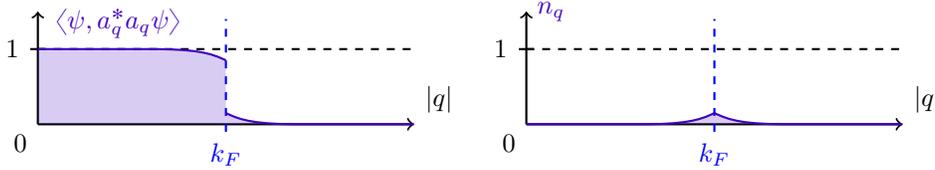
\begin{figure}[hbt]
	\centering
	\scalebox{1.0}{\begin{tikzpicture}

\draw[thick, ->] (0,0) -- ++(0,1.5);
\node[blue!75!red, anchor = south west] at (0.1,1) {\footnotesize $ \langle \psi, a_q^* a_q \psi \rangle $};
\draw[thick, ->] (0,0) -- ++(5,0) node[anchor = south west] {\footnotesize $ |q| $};
\node[anchor = north east] at (0,0) {\footnotesize $ 0 $};
\draw[thick] (0.1,1) -- ++(-0.2,0) node[anchor = east] {\footnotesize $ 1 $};
\draw[thick, dashed] (0,1) -- (5,1);
\draw[thick, blue, dashed] (2.5,-0.1) node[anchor = north] {\footnotesize $ k_F $} -- ++(0,1.5) ;

\draw[thick, blue!75!red] (0,1) -- (1.5,1) .. controls ++(0.6,0) and ++(-0.2,0.1) .. (2.5,0.85);
\draw[thick, blue!75!red] (2.5,0.15) .. controls ++(0.2,-0.1) and ++(-0.6,0) .. (3.5,0) -- (5,0);
\fill[blue!75!red, opacity = .2] (0,1) -- (1.5,1) .. controls ++(0.6,0) and ++(-0.2,0.1) .. (2.5,0.85) -- (2.5,0) -- (0,0);
\fill[blue!75!red, opacity = .2] (2.5,0.15) .. controls ++(0.2,-0.1) and ++(-0.6,0) .. (3.5,0) -- (2.5,0);

\end{tikzpicture}}
	\scalebox{1.0}{\begin{tikzpicture}

\draw[thick, ->] (0,0) -- ++(0,1.5);
\node[blue!75!red, anchor = west] at (0,1.5) {\footnotesize $ n_q $};
\draw[thick, ->] (0,0) -- ++(5,0) node[anchor = south west] {\footnotesize $ |q| $};
\node[anchor = north east] at (0,0) {\footnotesize $ 0 $};
\draw[thick] (0.1,1) -- ++(-0.2,0) node[anchor = east] {\footnotesize $ 1 $};
\draw[thick, dashed] (0,1) -- (5,1);
\draw[thick, blue, dashed] (2.5,-0.1) node[anchor = north] {\footnotesize $ k_F $} -- ++(0,1.5) ;

\draw[thick, blue!75!red] (0,0) -- (1.5,0) .. controls ++(0.6,0) and ++(-0.2,-0.1) .. (2.5,0.15);
\draw[thick, blue!75!red] (2.5,0.15) .. controls ++(0.2,-0.1) and ++(-0.6,0) .. (3.5,0) -- (5,0);
\fill[blue!75!red, opacity = .2] (1.5,0) .. controls ++(0.6,0) and ++(-0.2,-0.1) .. (2.5,0.15) -- (2.5,0) -- (1.5,0);
\fill[blue!75!red, opacity = .2] (2.5,0.15) .. controls ++(0.2,-0.1) and ++(-0.6,0) .. (3.5,0) -- (2.5,0);

\end{tikzpicture}}
	\caption{Left: Schematic depiction of the momentum distribution $ \langle \psi, a_q^* a_q \psi \rangle $, where the Fermi momentum $ k_{\F} $ is the radius of the Fermi ball $ B_{\F} = \overline{B_{k_{\F}}(0)} $.\\ Right: Excitation density $ n_q $ for the same trial state $ \psi $.}
	\label{fig:nq}
\end{figure}

Since physical systems often feature large particle numbers, it is of particular interest to derive mathematical statements in the limit $ N = \rho L^3 \to \infty $. Evidently, there are several choices of sequences in $ (\rho, L) $ for which $ N \to \infty $, where we will focus on the \textbf{mean-field limit} characterized by
\begin{equation}
	L := 2 \pi \;, \qquad
	\hbar := N^{-\frac 13} \;, \qquad
	\lambda := N^{-1} \;, \qquad
	\rho \to \infty \;.
\end{equation} 
We consider a sequence of trial states $ \psi_N = \psi $, first introduced in \cite{benedikter2020optimal} and close to the ground state. The quantity we are interested in is the \textbf{momentum distribution} $ q \mapsto \langle \psi, a_q^* a_q \psi \rangle \in [0, 1] $, with momentum $ q \in \ZZZ^3 $ and where $ a_q^*, a_q $ are the fermionic creation and annihilation operators defined below. Physically, $ \langle \psi, a_q^* a_q \psi \rangle $ describes the probability that a momentum mode $ q $ is occupied. It is well-known that for $ \lambda = 0 $, this probability is 1 if $ q $ is inside the Fermi ball $ B_{\F} := \overline{B_{k_{\F}}(0)} $ for some Fermi momentum $ k_{\F} $, and 0 outside $ B_{\F} $. The deviation from this profile for $ \lambda > 0 $ is called \textbf{excitation density}
\begin{equation}
\label{eq:nq}
	n_q
	:= \begin{cases}
		\langle \psi, a_q^* a_q \psi \rangle		\quad &\text{for } q \in B_{\F}^c \\
		1 - \langle \psi, a_q^* a_q \psi \rangle	\quad &\text{for } q \in B_{\F}
	\end{cases} \;,
\end{equation}
see Fig.~\ref{fig:nq}. $ n_q $ is expected to be small for small $ \lambda $, and it fully characterizes the momentum distribution. Our \textbf{main result}, Theorem \ref{thm:multicommutatorexact}, is now an exact formula for $ n_q $ in terms of so-called Friedrichs diagrams. Morally, it reads
\begin{equation}
	n_q = \sum_{\mathrm{diagrams}} \mathrm{value \; of \; the \; diagram} \;,
\end{equation}
with the sum being infinite but convergent. Friedrichs diagrams are a convenient tool to facilitate computationally intensive (anti-)commutator evaluations. They first appeared in \cite{friedrichs1965perturbation} and soon found a widespread application in constructive quantum field theory (CQFT) \cite{hepp1969theorie,glimm1968boson,glimm1973positivity,feldman1976wightman}. For further information on Friedrichs diagrams, we refer the interested reader to \cite{brooks2023friedrichs,derezinski2013mathematics}.\\
Theorem \ref{thm:multicommutatorexact} complements a recent result in \cite{benedikter2023momentum}, where an approximate excitation density $ n_q^{(\bos)} $ was derived using a bosonization technique, and proven to be the correct leading order expression for $ n_q $. We will demonstrate in Proposition \ref{prop:multicommutatorbos} how $ n_q^{(\bos)} $ arises from an intuitive restriction to certain ``bosonized diagrams'' as
\begin{equation}
	n_q^{(\bos)} = \sum_{\mathrm{bosonized \; diagrams}} \mathrm{value \; of \; the \; diagram} \;.
\end{equation}

Let us remark that a similar diagrammatic formalism was applied in the early physics literature for deriving the ground state energy of a Fermi gas with Coulomb interaction (Jellium) at high densities \cite{gell1957correlation,sawada1957correlation,daniel1960momentum}. In particular, \cite{daniel1960momentum} employs a Feynman--Hellmann argument to obtain formulas for the momentum distribution, which agree with the one implied by $ n_q^{(\bos)} $ for short-ranged interactions, see \cite[Appendix~B]{benedikter2023momentum}. However, the formalism in \cite{gell1957correlation,sawada1957correlation,daniel1960momentum} corresponds to Feynman diagrams rather than Friedrichs diagrams: Lines represent propagators rather than Kronecker deltas and vertices represent interactions in the interaction picture at distinct times.\\
Other results in the physics literature on the ground state energy of a 3d Fermi gas include \cite{bohm1957role,bohm1953collective,sawada1957correlation2,coldwell1960zero,coldwell1963erratum,lam1971correlation}. From \cite{lam1971correlation}, further formulas for the momentum distribution involving spin were derived by a Feynman--Hellmann argument in \cite{lam1971momentum}.\\

The trial state we use is of the form $ \psi = R T \Omega $ with $ R $ being a particle-hole transformation and $ T $ an ``almost bosonic'' Bogoliubov transformation, based on a patch construction, which we describe below in Sect.~\ref{sec:setting}. It was first used in \cite{benedikter2020optimal} to derive an upper bound on the ground state energy of a 3d mean-field Fermi gas, with\footnote{Here, $ \hat{V} $ denotes the Fourier transform of $ V $.} $ \supp \hat{V} $ being compact. The same construction was shortly afterwards used to provide a matching lower bound for weak interactions (compact $ \supp \hat{V} $ and $ \Vert \hat{V} \Vert_\infty $ small) \cite{benedikter2021correlation} and stronger interactions ($ \sum_k |k| \hat{V}(k) < \infty $) \cite{benedikter2023correlation}, as well as for a description of the dynamics generated by $ H_N $ with $ \supp \hat{V} $ compact \cite{benedikter2022bosonization}. Another almost bosonic  Bogoliubov transformation similar to $ T $, but not depending on a patch construction, was introduced in \cite{christiansen2021random}, allowing for an upper and lower bound on the ground state energy in case $ \sum_k |k| \hat{V}(k) < \infty $. This transformation was employed to determine the lower excitation spectrum \cite{christiansen2022effective} and proving an upper bound on the ground state energy \cite{christiansen2023gell}, both in case $ \sum_k \hat{V}(k)^2 < \infty $. This also covers the Coulomb case $ \hat{V}(k) = |k|^{-2} $ for $ k \neq 0 $ and $ \hat{V}(0) = 0 $, also called Jellium model. Note that all references above in this paragraph assume $ \hat{V} \ge 0 $.\\
Another way to send $ N \to \infty $ for a 3d Fermi gas \eqref{eq:HN} is given by the thermodynamic limit, where one fixes $ \rho $ and takes $ L \to \infty $. Here, in the dilute regime ($ \rho \ll 1 $), further ``almost bosonic'' Bogoliubov transformations have been introduced, which are based on solutions of the scattering equation. They allow for proving upper and lower bounds on the ground state energy for smooth $ V \ge 0 $ \cite{falconi2021dilute,giacomelli2022bogoliubov,giacomelli2023optimal}.\\
The proof of ground state energy formulas in the thermodynamic limit for large $ \rho $ remains a challenging open problem. Also, the rigorous establishment of momentum distribution formulas for the true ground state, rather than just some trial state $ \psi $, is an interesting open task.\\
Let us further remark that in the 1d Luttinger model \cite{luttinger1963exactly}, the exact momentum distribution for the ground state has been obtained long ago \cite{lieb1965exact}, as the model is exactly solvable.\\

The rest of this article is structured as follows. In Sect.~\ref{sec:setting} we introduce the mathematical notation that is needed to define the trial state $ \psi $ in \eqref{eq:psi}. Sect.~\ref{sec:momentumdist} contains our main result, Theorem \ref{thm:multicommutatorexact}, as well as Proposition \ref{prop:cont}, preceded by the minimal definitions needed to write down the diagrammatic contributions. The actual Friedrichs diagram formalism is introduced in Sect.~\ref{sec:friedrichsdiagrams}. It allows us to prove Theorem \ref{thm:multicommutatorexact} in Sect.~\ref{sec:multicomm}, where it also becomes clear how the diagrammatic contributions arise from actual diagrams. In Sect.~\ref{sec:bosonization}, we heuristically motivate why the largest contributions to $ n_q $ come from the ``bosonized'' diagrams in $ n_q^{(\bos)} $, and we finally prove Proposition \ref{prop:multicommutatorbos}.\\

\section{Bosonized Operators and Trial State}
\label{sec:setting}

We mostly adopt the setting of \cite{benedikter2023momentum}. It is well-known that in the non-interacting case, $ V = 0 $, one ground state of $ H_N $ is given by a Slater determinant, called Fermi ball state
\begin{equation}
\label{eq:planewaveslater}
	\psi_{\mathrm FB}(x_1,x_2,\ldots,x_N)
	:= \frac{1}{\sqrt{N!}} \det\left(\frac{1}{(2\pi)^{3/2}} e^{\imag k_j \cdot x_i}\right)_{j,i=1}^N \;,
\end{equation}
with disjoint momenta $ (k_j)_{j = 1}^N \subset \ZZZ^3 $ minimizing the kinetic energy $ \sum_{j = 1}^N \hbar^2 |k_j|^2 $. More precisely, the $ k_j $ occupy the Fermi ball
\begin{equation}
\label{eq:fermiball}
	B_{\F} := \{k \in \ZZZ^3 \; \vert \; |k| \le k_{\F} \} \quad \textnormal{for some Fermi momentum} \quad k_{\F} > 0\;,
\end{equation}
where in particular, we assume\footnote{That is, when taking $ N \to \infty $, we restrict to a sequence $ (N_n)_{n \in \NNN} $ such that $ N_n = | \overline{ B_{\kF^{(n)}}(0) } \cap \ZZZ^3 | $ for some $ (\kF^{(n)})_{n \in \NNN} \subset \RRR $} $ N = |B_{\F}| $.\\

For $ V \neq 0 $, the true ground state of $ H_N $ has a highly non-trivial structure. We will therefore approximate it by the ($ N $-dependent) trial state $ \psi = \psi_N $ introduced in \cite{benedikter2020optimal}. To define $ \psi $, it is convenient to work in the language of second quantization. To this end, we introduce the fermionic Fock space
\[
\cF := \bigoplus_{n = 0}^\infty L_{\a}^2(\TTT^{3n})\;,
\]
and the orthonormal plane wave basis $ (f_q)_{q \in \ZZZ^3} \subset L^2(\TTT^3) $ with $ f_q(x) := (2 \pi)^{-\frac 32} e^{\imag q \cdot x} $, $ q \in \ZZZ^3 $ and the corresponding creation and annihilation operators 
\[
	a_q^*, a_q : \cF \to \cF \;, \qquad	
	a_q^* := a^*(f_q) \;, \qquad
	a_q := a(f_q) \;,
\]
which satisfy the canonical anticommutation relations (CAR)
\begin{equation}
\label{eq:CAR}
	\{a_q, a_{q'}^*\} = \delta_{q, q'} \;, \qquad
	\{a_q, a_{q'}\} = \{a_q^*, a_{q'}^*\} = 0 \qquad
	\text{for all } q, q' \in \ZZZ^3 \;.
\end{equation}
The vacuum vector $ \Omega \in \cF $ is given by $ \Omega := (1, 0, 0, \ldots) $ and satisfies $ a_q \Omega = 0 $ for all $q \in \ZZZ^3 $. Further, the Fermi ball state $ \psi_{\mathrm{FB}} $ \eqref{eq:planewaveslater} can now conveniently be written as $ \psi_{\mathrm FB} = R \Omega $ with $ R = R^*: \cF \to \cF $ being a unitary \textbf{particle--hole transformation} defined via
\begin{equation}
	R^* a_q^* R := \begin{cases}
		a_q^* \quad &\text{if } q \in B_{\F}^c\\
		a_q \quad &\text{if } q \in B_{\F}
	\end{cases} \;.
\end{equation}
By contrast, our \textbf{trial state} is of the form
\begin{equation}
\label{eq:psi}
	\psi := R T \Omega\;, \qquad \psi \in L_{\a}^2(\TTT^{3N}) \subset \cF \;,
\end{equation}
where $ T: \cF \to \cF $ is a unitary ``almost bosonic Bogoliubov transformation'', which we define in the following.\\

\begin{figure}[hbt]
	\centering
	\includegraphics[width=4.5cm]{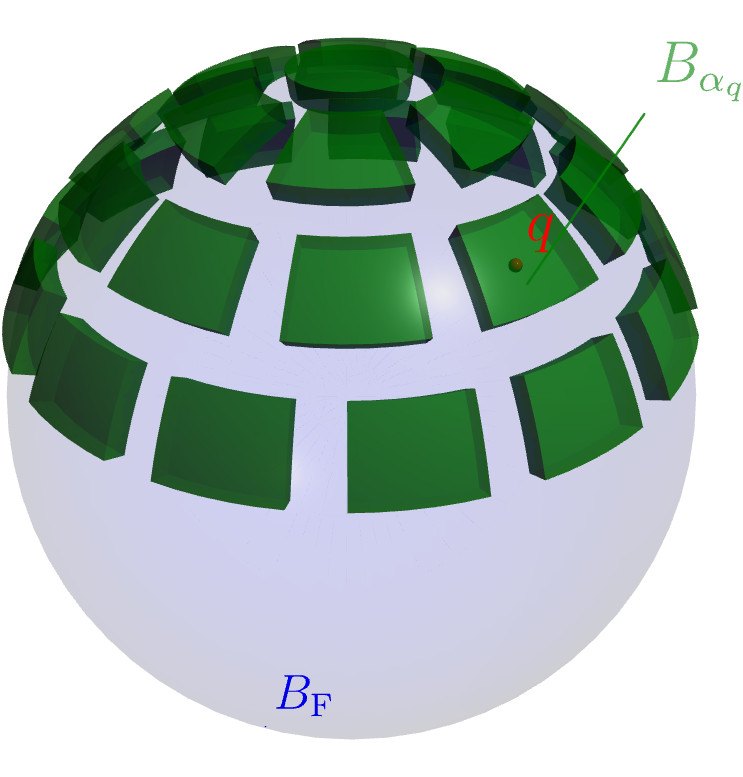}
 	\caption{Patches on the Fermi ball in momentum space.}
 \label{fig:}
	\label{fig:patches3d}
\end{figure}

The first step in the definition of $ T $ consists of constructing $ M $ \textbf{patches} $ B_1, \ldots, B_M \subset \RRR^3 $ around the Fermi surface $ \partial \BF $, see Fig.~\ref{fig:patches3d}. Details of the construction are given in \cite{benedikter2020optimal,benedikter2021correlation} and involve some fixed constant $ R > 0 $. Each patch covers approximately the same surface area of $ \frac{4 \pi k_{\F}^2}{M} $, has a thickness of $ 2 R $, and is separated from the neighboring patches by a corridor of width $ > R $. Its edges are assumed to be regular in the sense that
\[
	\diam(B_\alpha)
	\le C N^{\frac{1}{3}} M^{-\frac{1}{2}} \;,
\]
and the patch number $ M $ is chosen such that $ N^{2\delta} \ll M \ll N^{\frac{2}{3}- 2 \delta} $ for some fixed $ 0 < \delta < \frac{1}{6} $. More precisely, in~\cite{benedikter2023momentum}, we fix $ M = N^{\frac 13} $ and $ \delta = \frac{1}{12} $.\\
Denoting the center of a patch $ B_\alpha $ by $ \omega_\alpha \in \RRR^3 $ with corresponding unit vector $ \hat{\omega}_\alpha := \frac{\omega_\alpha}{|\omega_\alpha|} $, we define for every fixed $ k \in \ZZZ^3 $ the index sets
\begin{equation}
\label{eq:cIk}
\begin{aligned}
	\cI_{k}^+ & := \left\{ \alpha \in \{1, \ldots, M\} \; \big\vert \; k \cdot \hat{\omega}_\alpha \ge N^{-\delta}\right\} \;, \\
	\cI_{k}^- & := \left\{ \alpha \in \{1, \ldots, M\} \; \big\vert \; k \cdot \hat{\omega}_\alpha \le -N^{-\delta}\right\} \;, \\
	\cI_{k} & := \cI_{k}^+ \cup \cI_{k}^- \;.
\end{aligned}
\end{equation}
The patch construction is assumed to be symmetric under replacement $ k \mapsto -k $, so we have $ |\cI_{k}^+| = |\cI_{k}^-| \le \frac M2 $. The symmetry allows us to restrict our focus to momenta $ k $ within the northern half-sphere
\begin{equation}
\label{eq:Hnor}
	H^{\nor} := \big\{ k \in \RRR^3 \; \big\vert \; k_3 > 0 \text{ or } (k_3 = 0 \text{ and } k_2 > 0)
	\text{ or } (k_3 = k_2 = 0 \text{ and } k_1 > 0) \big\} \;.
\end{equation}
Let us further adopt the following conventions: All momenta $ k, q, p, h $ are assumed to be elements of $ \ZZZ^3 $. For $ p \in \ZZZ^3 $ (``particle''), the condition $ p \in B_{\F}^c \cap B_\alpha $ is abbreviated as $ p : \alpha $ (read as ``$ p $ is compatible with $ B_\alpha $'') and for $ h \in \ZZZ^3 $ (``hole''), the condition $ h \in B_{\F} \cap B_\alpha $ is abbreviated as $ h : \alpha $. Further, we introduce the notation
\[
	\pm k := \begin{cases}
		+k \quad \text{if } \alpha \in \cI_{k}^+\\
		-k \quad \text{if } \alpha \in \cI_{k}^-\\
	\end{cases}
	\; \text{and} \quad
	\mp k := \begin{cases}
		-k \quad \text{if } \alpha \in \cI_{k}^+\\
		+k \quad \text{if } \alpha \in \cI_{k}^-\\
	\end{cases} \;.
\]
Given a momentum transfer
\begin{equation}
\label{eq:Gammanor}
	k \in \Gamma^{\nor} := H^{\nor} \cap \ZZZ^3 \cap B_R(0) \;,
\end{equation}
the number of particle--hole pairs with $ p = h \pm k $ in patch $ B_\alpha \in \cI_k $ is then
\begin{equation}
\label{eq:n}
	n_{\alpha, k}^2 
	:= \sum_{p, h : \alpha} \delta_{p, h \pm k} 
	= \sum_{\substack{p: p \in B_{\F}^c \cap B_\alpha \\ p \mp k \in B_{\F} \cap B_\alpha}} 1 \;.
\end{equation}
We now define for $ k \in \Gamma^{\nor} $ and $ B_\alpha \in \cI_k $ the \textbf{almost bosonic creation operator}
\begin{equation}
\label{eq:cstar}
	c^*_\alpha(k)
	:= \frac{1}{n_{\alpha, k}} \sum_{p, h: \alpha} \delta_{p, h \pm k} a_p^* a_h^*
	= \frac{1}{n_{\alpha, k}} \sum_{\substack{p: p \in B_{\F}^c \cap B_\alpha \\ p \mp k \in B_{\F} \cap B_\alpha}} a_p^* a_{p \mp k}^* \;,
\end{equation}
with its adjoint being $ c_\alpha(k) $. The normalization factor $ n_{\alpha, k} $ is precisely chosen such that $ c^*, c $ satisfy the \textbf{approximate canonical commutation relations} (CCR) \cite{benedikter2023momentum,benedikter2020optimal,benedikter2021correlation}
\begin{equation}
\label{eq:approxCCR}
	[c_\alpha(k), c^*_\beta(\ell)] = \begin{cases}
		0 \quad &\text{if } \alpha \neq \beta\\
		\delta_{k, \ell} + \cE_\alpha(k, \ell) \quad &\text{if } \alpha = \beta
	\end{cases} \;,
\end{equation}
with commutation error
\begin{equation}
\label{eq:cEalphaql}
	\cE_\alpha(k, \ell) :=
	- \sum_{p, h_1, h_2 : \alpha} \frac{\delta_{h_1, p \mp k} \delta_{h_2, p \mp \ell}}{n_{\alpha, k} n_{\alpha, \ell}} a_{h_2}^* a_{h_1}
	- \sum_{p_1, p_2, h : \alpha} \frac{\delta_{h, p_1 \mp k} \delta_{h, p_2 \mp \ell}}{n_{\alpha, k} n_{\alpha, \ell}} a_{p_2}^* a_{p_1}\;.
\end{equation}
The \textbf{almost bosonic Bogoliubov transformation} $ T: \cF \to \cF $ generating the trial state $ \psi = R T \Omega $ is then of the form
\begin{equation}
\label{eq:T}
	T := e^{-S}\;, \qquad
	S := -\frac{1}{2} \sum_{k \in \Gamma^{\nor}} \sum_{\alpha, \beta \in \cI_{k}} K(k)_{\alpha, \beta} \big( c^*_\alpha(k) c^*_\beta(k) - \mathrm{h.c.} \big) \;,
\end{equation}
with $ K(k) \in \RRR^{|\cI_{k}| \times |\cI_{k}|} $ being a symmetric matrix defined below. So $ S^* = -S $ and
\begin{equation}
\label{eq:S+rewriting}
\begin{aligned}
	S = &S_+ + S_- \;, \quad \text{where} \quad S_- = - (S_+)^* \quad \text{and}\\ 
	S_+ := &- \frac 12 \sum_{k \in \Gamma^{\nor}}
	\sum_{\alpha, \alpha' \in \cI_{k}} K(k)_{\alpha, \alpha'}
	\sum_{\substack{p, h : \alpha \\ p', h' : \alpha'}}
	\frac{ \delta_{p, h \pm k} \delta_{p', h' \pm k} }{n_{\alpha, k} n_{\alpha', k}}
	a_{p}^* a_{h}^* a_{p'}^* a_{h'}^* \;.
\end{aligned}
\end{equation}
The motivation behind the definition of $ T $ is to ``almost-diagonalize'' some effective quadratic bosonic Hamiltonian, see \cite{benedikter2020optimal,benedikter2021correlation}. This requires the following choice of $ k $-dependent matrices:
\begin{align*}
	K &:= \log |S_1^T|\;, \tagg{eq:K}\\
	S_1 &:= (D + W - \widetilde{W})^{\frac{1}{2}} E^{-\frac{1}{2}}\;, \\
	E &:= \left( (D + W - \widetilde{W})^{\frac{1}{2}} (D + W + \widetilde{W}) (D + W - \widetilde{W})^{\frac{1}{2}} \right)^{\frac{1}{2}} \;,
\end{align*}
with the $\RRR^{|\cI_{k}| \times |\cI_{k}|} $ symmetric block matrices
\[
	D = \begin{pmatrix}
		\; d \; & \; 0 \; \\
		\; 0 \; & \; d \; \\
	\end{pmatrix}\;, \qquad
	W = \begin{pmatrix}
		\; b \; & \; 0 \; \\
		\; 0 \; & \; b \; \\
	\end{pmatrix}\;, \qquad
	\widetilde{W} = \begin{pmatrix}
		\; 0 \; & \; b \; \\
		\; b \; & \; 0 \; \\
	\end{pmatrix}\;,
\]
where $ d, b \in \RRR^{|\cI_{k}^+| \times |\cI_{k}^+|} $ are given by
\begin{equation}
\label{eq:db}
	d := \sum_{\alpha \in \cI_{k}^+} |\hat{k} \cdot \hat{\omega}_\alpha| \; |\alpha \rangle \langle \alpha |\;, \qquad
	b := \sum_{\alpha, \beta \in \cI_{k}^+} \frac{\hat{V}(k)}{2 \hbar \kappa N |k|} n_{\alpha, k} n_{\beta, k} \; |\alpha \rangle \langle \beta |\;.
\end{equation}
Here, $ | \alpha \rangle \in \RRR^{|\cI_{k}^+|} $ is the $ \alpha $-th canonical basis vector, we have $ \hat{k} := k/|k| $ and $ \kappa := k_{\F} N^{-\frac 13} \approx \left( \frac{3}{4 \pi} \right)^{\frac 13} $. This concludes the construction of the trial state $ \psi $.\\

\section{Momentum Distribution}
\label{sec:momentumdist}

In \cite[Thm.~3.1]{benedikter2023momentum}, it was proved that for $ \hat{V} \ge 0 $ compactly supported, the excitation density $ n_q $ \eqref{eq:nq} for $ q \in \alpha_q $ with some $ 1 \le \alpha_q \le M $ is approximately given by some \textbf{bosonized excitation density} \cite[(5.7)]{benedikter2023momentum} of the form
\begin{equation}
\label{eq:nqb}
	n_q^{(\bos)}
	:= \frac{1}{2} \sum_{k \in \tilde{\cC}^q \cap \ZZZ^3} \frac{1}{n_{\alpha_q, k}^2} \big( \cosh(2 K(k)) - 1 \big)_{\alpha_q, \alpha_q}\;,
\end{equation}
in the sense that for fixed $ V $ and most $ q $,
\begin{equation}
	|n_q - n_q^{(\bos)}|
	\le C_\varepsilon N^{-\frac 23 - \frac{1}{12}} \;,
\end{equation}
with $ C_\varepsilon > 0 $ depending only on $ \varepsilon > 0 $ and not on $ N $. Here $ \tilde{\cC}^q $ \cite[(3.1)]{benedikter2023momentum} is defined such that
\begin{equation}
\label{eq:cCq}
	\tilde{\cC}^q \cap \ZZZ^3 = \begin{cases}
		\big\{ k \in \Gamma^{\nor} \; \big\vert \;
			\alpha_q \in \cI_k \text{ and }
			q \mp k : \alpha_q	
		\big\} \quad &\text{if } q \in B_{\F}^c \\
		\big\{ k \in \Gamma^{\nor} \; \big\vert \;
			\alpha_q \in \cI_k \text{ and }
			q \pm k : \alpha_q	
		\big\} \quad &\text{if } q \in B_{\F}
	\end{cases} \;.
\end{equation}
The approximation $ n_q^{(\bos)} $ arises when evaluating the multicommutator series
\begin{equation}
\label{eq:BCH}
	n_q
	= \langle \Omega, T^* a_q^* a_q T \Omega \rangle
	= \langle \Omega, e^S a_q^* a_q e^{-S} \Omega \rangle
	= \sum_{n = 0}^\infty \frac{1}{n!} \langle \Omega, \ad^n_S (a_q^* a_q) \Omega \rangle \;,
\end{equation}
with $ \ad^n_A(B) = [A, \ldots, [A, [A,B]] \ldots ] $ being the $ n $-fold multicommutator.\\
We will now provide an exact formula \eqref{eq:multicommutatorexact} for $ n_q $, which requires introducing some notation. It turns out that only even $ n $ render contributions to $ n_q $ in \eqref{eq:BCH}. For such $ n $, we denote the momenta of the $ n $ involved $ S $-operators, see \eqref{eq:S+rewriting}, by $ \{ p_j, p_j', h_j, h_j' \}_{j = 1}^n $. For later convenience we write
\begin{equation}
\label{eq:aqstaraqsplit}
	a_q^* a_q = \begin{cases}
		\sum_{p_0, p_0'} \delta_{q, p_0} \delta_{q, p_0'} \; a_{p_0}^* a_{p_0'} \quad &\text{for } q \in B_{\F}^c \\
		\sum_{h_0, h_0'} \delta_{q, h_0} \delta_{q, h_0'} \; a_{h_0}^* a_{h_0'} \quad &\text{for } q \in B_{\F} \\
	\end{cases} \;,
\end{equation}
which allows to write the involved momentum indices in the multicommutator as
\begin{equation}
\label{eq:bPbH}
\begin{aligned}
	\bP &:= (p_0, p_1, \ldots , p_n) \;, \qquad &\bP' &:= (p_0', p_1',\ldots , p_n') \;,\\
	\bH &:= (h_1, \ldots , h_n) \;, \qquad &\bH' &:= (h_1', \ldots , h_n') \;,\\
\end{aligned}
\end{equation}
for $ q \in B_{\F}^c $. In case $ q \in B_{\F} $, we analogously include $ h_0 $ and $ h_0' $ into $ \bH $ and $ \bH' $. We will further split each commutator with $ S = S_+ + S_- $ \eqref{eq:S+rewriting} into one commutator with $ S_+ $ and one with $ S_- $, where the choices between $ S_+ $ or $ S_- $ are tracked by a map
\begin{equation}
\label{eq:Xin}
	\xi \in \Xi_n := \Big\{ \xi : \{1, \ldots, n\} \mapsto \{1,-1\} \; \Big\vert \; \sum_{j = 1}^n \xi(j) = 0 \Big\} \;.
\end{equation}
Here, $ \xi(j) = 1 $ means that $ S_+ $ and $ \xi(j) = -1 $ that $ S_- $ is chosen for the $ j $-th commutator. According to $ \xi \in \Xi_n $, we may split the momentum indices into those belonging to creation and annihilation parts. In case $ q \notin B_{\F} $,
\begin{equation}
\label{eq:bPbHpartitions}
\begin{aligned}
	\bP &= \bP_+ \cup \bP_- \;, \;
		&\bP_+ &:= p_0 \cup (p_j : \xi(j) = 1) \;, \;
		&\bP_- &:= (p_j : \xi(j) = 0) \;,\\
	\bH &= \bH_+ \cup \bH_- \;, \;
		&\bH_+ &:= (h_j : \xi(j) = 1) \;, \;
		&\bH_- &:= (h_j : \xi(j) = 0) \;,\\
	\bP' &= \bP'_+ \cup \bP'_- \;, \; 
		&\bP'_+ &:=(p_j' : \xi(j) = 1) \;, \; 
		&\bP'_- &:= p_0' \cup (p_j' : \xi(j) = 0) \;,\\
	\bH' &= \bH'_+ \cup \bH'_- \;, \; 
		&\bH'_+ &:= (h_j' : \xi(j) = 1) \;, \; 
		&\bH'_- &:= (h_j' : \xi(j) = 0) \;.\\
\end{aligned}
\end{equation}
In case $ q \in B_{\F} $, we analogously include $ h_0 $ in $ \bH_+ $ and $ h_0' $ in $ \bH'_- $.\\
The contractions will now each be between one creation- ($ + $) and one annihilation momentum index ($ - $). We track them by two bijective maps
\begin{equation}
\label{eq:pippih}
	\pi_p : \bP_- \cup \bP'_- \mapsto \bP_+ \cup \bP'_+ \;, \qquad
	\pi_h : \bH_- \cup \bH'_- \mapsto \bH_+ \cup \bH'_+ \;.
\end{equation}
All contractions are subject to the constraint that for each $ S_\pm $-operator, at least one momentum index is contracted to an ``earlier'' index:
\begin{equation}
\label{eq:piconstraint}
\begin{aligned}
	&\forall j \in \{1, \ldots n \} \; \exists \ell \in \{0, \ldots , j-1\} :\\
	&\qquad \begin{cases}
		\{ \pi_p(p_{\ell}), \pi_p(p_{\ell}'), \pi_h(h_{\ell}), \pi_h(h_{\ell}') \} \cap \{p_j, p_j', h_j, h_j'\} \neq \emptyset \qquad \text{if } \xi(j) = 1 \\
		\{ \pi_p(p_j), \pi_p(p_j'), \pi_h(h_j), \pi_h(h_j') \} \cap \{p_{\ell}, p_{\ell}', h_{\ell}, h_{\ell}'\} \neq \emptyset \qquad \text{if } \xi(j) = -1 \\
	\end{cases} \;.
\end{aligned}
\end{equation}
The emergence of this constraint will become apparent later in Sect.~\ref{sec:multicomm}. We then denote the set of \textbf{admissible contraction choices} by
\begin{equation}
\label{eq:Pin}
	\Pi_n^{(\xi)} := \{ (\pi_p, \pi_h) \; \vert \; \eqref{eq:piconstraint} \text{ holds} \} \;.
\end{equation}
To each contraction choice, we will now associate a sign factor $ \sgn(\xi, \pi_p, \pi_h) \in \{1, -1\} $. In order to define this sign factor, let us introduce the sets of creation- and annihilation associated momenta
\begin{equation}
\label{eq:bQ}
	\bQ_+ := \bP_+ \cup \bP'_+ \cup \bH_+ \cup \bH'_+ \;, \qquad
	\bQ_- := \bP_- \cup \bP'_- \cup \bH_- \cup \bH'_- \;,
\end{equation}
as well as two ordering relations $ < $ on $ \bQ_- $ and on $ \bQ_+ $, respectively, defined by
\begin{equation}
\label{eq:orderingrelation}
	\ell < j \; \Rightarrow \; p_{\ell}, h_{\ell}, p'_{\ell}, h'_{\ell} < p_j, h_j, p'_j, h'_j \quad \text{and} \quad p_j < h_j < p'_j < h'_j \;.
\end{equation}
To each contraction choice, we associate a sign factor
\begin{equation}
\label{eq:sgnxipippih}
	\sgn(\xi, \pi_p, \pi_h) := \prod_{q_- \in \bQ_-} \prod_{\substack{q_-' \in \bQ_- \\ q_-' > q_-, \; \pi_{\sharp}(q_-') < \pi_{\sharp}(q_-) }} (-1) \;,
\end{equation}
with $ \sharp \in \{p, h\} $, according to whether $ q_-, q_-' $ are particle or hole momenta. Finally, let us abbreviate the momentum transfer and patch indices involved in the $ n $ operators $ S $ by
\begin{equation}
\label{eq:bKbalpha}
	\bK := (k_1, \ldots, k_n) \;, \qquad
	\balpha := (\alpha_1, \ldots, \alpha_n) \;, \qquad
	\balpha' := (\alpha_1', \ldots, \alpha_n') \;.
\end{equation}
We are now in the position to write down the exact formula for $ n_q $.

\begin{theorem}[Main result, exact excitation density]
Let $ q \in \ZZZ^3 $. If $ q \in B_{\F}^c $, then the excitation density in the trial state $ \psi = R T \Omega $ \eqref{eq:psi} from \cite{benedikter2020optimal} is given by
\begin{equation}
\label{eq:multicommutatorexact}
\begin{aligned}
	n_q = &\sum_{\substack{n = 2 \\ n : \mathrm{even}}}^\infty \frac{1}{2^n n!} \sum_{\bK} \sum_{\balpha, \balpha'} \sum_{\substack{\bP, \bP' \\ \bH, \bH'}} 
	\left( \prod_{j = 1}^n \frac{\delta_{p_j, h_j \pm k_j} \delta_{p_j', h_j' \pm k_j}}{n_{\alpha_j, k_j} n_{\alpha_j', k_j}} K(k_j)_{\alpha_j, \alpha_j'} \right) \times\\
	&\quad \times
	\sum_{\xi \in \Xi_n} \sum_{(\pi_p, \pi_h) \in \Pi_n^{(\xi)}}
	\left( \prod_{p \in \bP_- \cup \bP'_-} \delta_{p, \pi_p(p)} \right)
	\left( \prod_{h \in \bH_- \cup \bH'_-} \delta_{h, \pi_h(h)} \right)
	\delta_{q, p_0} \delta_{q, p_0'} \sgn(\xi, \pi_p, \pi_h)\;,
\end{aligned}
\end{equation}
where the sum in $ \bK $ runs over $ (\Gamma^{\nor})^n \subset \ZZZ^{3n} $ (see \eqref{eq:Gammanor}), the sums in $ \balpha, \balpha' $ are such that $ \alpha_j, \alpha_j' \in \cI_{k_j} $ (see \eqref{eq:cIk}), and the sums over $ \bP, \bP', \bH, \bH' $ are such that $ p_j, h_j : \alpha_j $ and $ p_j', h_j' : \alpha_j' $, that is,
\begin{equation}
\label{eq:phcondition}
\begin{aligned}
		p_j \in B_{\F}^c \cap B_{\alpha_j} \quad \mathrm{and} \quad
		&&h_j \in B_{\F} \cap B_{\alpha_j} \quad \forall j \in \{1, \ldots, n \} \;, \\
		p_j' \in B_{\F}^c \cap B_{\alpha_j'} \quad \mathrm{and} \quad
		&&h_j' \in B_{\F} \cap B_{\alpha_j'} \quad \forall j \in \{1, \ldots, n \} \;.\\
\end{aligned}
\end{equation}
In case $ q \in B_{\F} $, \eqref{eq:multicommutatorexact} remains valid after a replacement of $ \delta_{q, p_0} \delta_{q, p_0'} $ by $ \delta_{q, h_0} \delta_{q, h_0'} $.
\label{thm:multicommutatorexact}
\end{theorem}

The proof is given in Sect.~\ref{sec:multicomm} using Friedrichs diagrams. There, it will also become clear how the rather involved term in \eqref{eq:multicommutatorexact} arises from a step-by-step back-translation of Friedrichs diagrams into a mathematical expression.\\

\begin{remark}
\textit{Convergence of the diagrammatic expansion.} The expansion \eqref{eq:multicommutatorexact} indeed converges: We obtain it by a separate evaluation of each order $ n \in \NNN $ in the multicommutator expansion \eqref{eq:BCH}. Since the number of terms (i.e., diagrams) per order $ n $ is finite, the expansion \eqref{eq:multicommutatorexact} converges if and only if \eqref{eq:BCH} converges. Now, $ \Vert a_q^* a_q \Vert = 1 $, so if we can show that $ S $ is bounded, then we obtain absolute convergence of the commutator series \eqref{eq:BCH} as
\begin{equation}
	\sum_{n = 0}^\infty \frac{|\langle \Omega, \ad^n_S(a_q^* a_q) \Omega \rangle|}{n!}
	\le \sum_{n = 0}^\infty \frac{\Vert \ad^n_S(a_q^* a_q) \Vert}{n!}
	\le \sum_{n = 0}^\infty \frac{\Vert 2 S \Vert^n \Vert a_q^* a_q \Vert}{n!}
	= e^{\Vert 2 S \Vert} \;.
\end{equation}
In fact, recalling \eqref{eq:T} we have
\begin{equation}
	\Vert S \Vert
	\le \sum_{k \in \Gamma^{\nor}} \sum_{\alpha, \beta \in \cI_k}
		|K(k)_{\alpha, \beta}|
		\Vert c^\sharp_\alpha(k) \Vert
		\Vert c^\sharp_\beta(k) \Vert \;.
\end{equation}
With \cite[Lemma~7.1]{benedikter2023correlation}, we estimate $ |K(k)_{\alpha, \beta}| \le C M^{-1} \hat{V}(k) $. Further, by \eqref{eq:cstar},
\begin{equation}
	\Vert c^*_\alpha(k) \Vert
	\le \frac{1}{n_{\alpha, k}} \sum_{\substack{p: p \in B_{\F}^c \cap B_\alpha \\ p \mp k \in B_{\F} \cap B_\alpha}} \Vert a_p^* a_{p \mp k}^* \Vert
	\le \frac{1}{n_{\alpha, k}} \sum_{\substack{p: p \in B_{\F}^c \cap B_\alpha \\ p \mp k \in B_{\F} \cap B_\alpha}} 1
	= n_{\alpha, k} \;,
\end{equation}
and the same bound holds for $ \Vert c_\alpha(k) \Vert $. With $ n_{\alpha, k}^2 \le C N^{\frac 23} M^{-1} $, which follows form the patch construction, and using that the sums $ \sum_\alpha, \sum_\beta $ run over $ \le M $ elements, we conclude
\begin{equation}
\label{eq:Snorm}
	\Vert 2 S \Vert
	\le C M^{-1} \sum_{k \in \Gamma^{\nor}} \hat{V}(k) \sum_{\alpha, \beta \in \cI_k} n_{\alpha, k} n_{\beta, k} 
	\le C N^{\frac 23} \sum_{k \in \Gamma^{\nor}} \hat{V}(k) 
	< \infty \;,
\end{equation}
since $ \Gamma^{\nor} $ comprises finitely many lattice points.\\
\end{remark}

\begin{remark}
\textit{Allowed potentials and scaling limits.} In Theorem \ref{thm:multicommutatorexact}, we did not specify any conditions on $ V $. Indeed, our result holds for \textit{any potential} $ V $, provided that $ \hat{V}: \ZZZ^3 \to \RRR $ exists. This may first seem somewhat surprising and is owed to the particular choice of the trial state $ \psi $, involving finite sums $ \sum_{k \in \Gamma^{\nor}} $ in $ S $ \eqref{eq:T}, and $ \sum_p $ in $ c^\sharp_\alpha(k) $ \eqref{eq:cstar}. However, if the potential $ \hat{V}(k) $ does not satisfy the requirements of \cite{benedikter2023correlation} that $ \sum_{k \in \ZZZ^3} \hat{V}(k) |k| < \infty $ and $ \hat{V} \ge 0 $, then there is no guarantee that $ \psi $ is a good approximation of the ground state (in terms of energy). So while Theorem \ref{thm:multicommutatorexact} is still correct, its physical significance then still has to be demonstrated.\\
Likewise, we could pick any side length $ L $ of the torus and analogously construct a trial state $ \psi $ using the refined momentum lattice $ \frac{2 \pi}{L} \ZZZ^3 $. Then, sums will still be finite and Theorem \ref{thm:multicommutatorexact} is still valid. In particular, Theorem \ref{thm:multicommutatorexact} holds for any element of a sequence of trial states $ (\psi_N)_{N \in \NNN} $, constructed as above, in \textit{any scaling limit}, including the thermodynamic limit at high density $ \rho $. However, there is no guarantee that these $ \psi_N $ will be good approximations for the ground state.\\
\end{remark}

\begin{remark}
\textit{Orders of the contributing diagrams.} As we explain in Sect.~\ref{subsec:heuristics}, the contractions $ \delta_{p, \pi_p(p)} $ and $ \delta_{h, \pi_h(h)} $ eliminate certain sums over $ \alpha_j, \alpha_j' $ by setting patch indices equal. As each sum runs over $ \sim M $ elements, we expect the diagrams, indexed by $ (\xi, \pi_p, \pi_h) $, to be of different orders, depending on how many sums survive. More precisely, we expect \eqref{eq:multicommutatorexact} to result in an expansion of the form
\begin{equation}
\label{eq:expansionM}
	n_q = f_0 + f_1 M^{-1} + f_2 M^{-2} + f_3 M^{-3} + \ldots \;,
\end{equation}
where the coefficients $ f_j \in \RRR, \; j \in \NNN $ have identical scaling in $ N $. The leading order is expected to be $ f_0 = n_q^{(\bos)} $, which scales like $ \sim N^{- \frac 23} $ for ``most'' $ q $, as explained in \cite[Remark~2]{benedikter2023momentum}. For the choice $ M = \frac 13 $ and $ \delta = \frac{1}{12} $ as above, the optimal relative error of the approximation $ n_q \approx n_q^{(\bos)} $ is thus expected to be $ M^{-1} = N^{-\frac{10}{27}} $, which is much smaller than the existing relative error bound $ N^{-\frac{2}{27}} $.\\
We believe that an expansion like \eqref{eq:expansionM} can be made rigorous, once suitable combinatorial bounds on the number of occurring diagrams have been derived.\\
\end{remark}

We are able to interpret the bosonized excitation density $ n_q^{(\bos)} \approx n_q $ obtained in \cite[Thm.~3.1]{benedikter2023momentum} as coming from a subset of bosonized Friedrichs diagrams, which we explain in Sect.~\ref{sec:bosonization}. The subset is characterized by a restriction on $ (\pi_p, \pi_h) $. First, we require that the two $ c^\sharp_\alpha(k) $-operators, whose momenta are contracted to $ a_q^* a_q $, have their second momenta directly contracted to each other. Second, for all other $ c_\alpha(k) $-operators, both momenta must be contracted at the same time to the same $ c^*_\alpha(k) $-operator. So in case $ q \in B_{\F}^c $, with $ \sharp_1, \sharp_2 \in \{ \cdot, '\} $ (so $ p^{\sharp_1} $ is either $ p $ or $ p' $), the \textbf{bosonization constraint} reads
\begin{equation}
\label{eq:piconstraintstrict}
\begin{aligned}
	\pi_p^{-1}(p_0) = p_j^{\sharp_1} \; \text{ and } \; \pi_p(p_0') = p_{\ell}^{\sharp_2} \quad &\Rightarrow \quad
	\pi_h(h_j^{\sharp_1}) = h_{\ell}^{\sharp_2} \\
	\forall j, \ell \ge 1: \qquad \pi_p(p_j^{\sharp_1}) = p_{\ell}^{\sharp_2} \quad &\Rightarrow \quad
	\pi_h(h_j^{\sharp_1}) = h_{\ell}^{\sharp_2} \;.
\end{aligned}
\end{equation}
The diagrammatic interpretation of this constraint is explained in Sect.~\ref{subsec:heuristics} and depicted in Fig.~\ref{fig:Friedrichs_loop}. The restricted set of \textbf{bosonized contraction choices} is then
\begin{equation}
\label{eq:Pinb}
	\Pi_{n, (\bos)}^{(\xi)} := \big\{ (\pi_p, \pi_h) \in \Pi_n^{(\xi)} \; \big\vert \; \eqref{eq:piconstraintstrict} \text{ holds} \big\} \;.
\end{equation}

\begin{proposition}[Bosonized excitation density]
The bosonized excitation density $ n_q^{(\bos)} $ \eqref{eq:nqb} from \cite{benedikter2023momentum} amounts to a restriction of \eqref{eq:multicommutatorexact} to bosonized diagrams, given by replacing $ \Pi_n^{(\xi)} $ with $ \Pi_{n, (\bos)}^{(\xi)} $. That is, for $ q \in B_{\F}^c $,
\begin{equation}
\label{eq:multicommutatorbos}
\begin{aligned}
	n_q^{(\bos)} = &\sum_{\substack{n = 2 \\ n : \mathrm{even}}}^\infty \frac{1}{2^n n!} \sum_{\bK} \sum_{\balpha, \balpha'} \sum_{\substack{\bP, \bP' \\ \bH, \bH'}} 
	\left( \prod_{j = 1}^n \frac{\delta_{p_j, h_j \pm k_j} \delta_{p_j', h_j' \pm k_j}}{n_{\alpha_j, k_j} n_{\alpha_j', k_j}} K(k_j)_{\alpha_j, \alpha_j'} \right) \times\\
	&\times
	\sum_{\xi \in \Xi_n} \sum_{(\pi_p, \pi_h) \in \Pi_{n, (\bos)}^{(\xi)}}
	\left( \prod_{p \in \bP_- \cup \bP'_-} \delta_{p, \pi_p(p)} \right)
	\left( \prod_{h \in \bH_- \cup \bH'_-} \delta_{h, \pi_h(h)} \right)
	\delta_{q, p_0} \delta_{q, p_0'} \sgn(\xi, \pi_p, \pi_h)\;,
\end{aligned}
\end{equation}
and the same holds for $ q \in B_{\F} $ after a replacement of $ \delta_{q, p_0} \delta_{q, p_0'} $ by $ \delta_{q, h_0} \delta_{q, h_0'} $.
\label{prop:multicommutatorbos}
\end{proposition}

We illustrate and explain the meaning of \eqref{eq:multicommutatorbos} in Sect.~\ref{subsec:heuristics} and give the diagrammatic proof of Proposition~\ref{prop:multicommutatorbos} in Sect.~\ref{subsec:bosonizationevaluation}.\\

\section{Friedrichs Diagram Formalism}
\label{sec:friedrichsdiagrams}

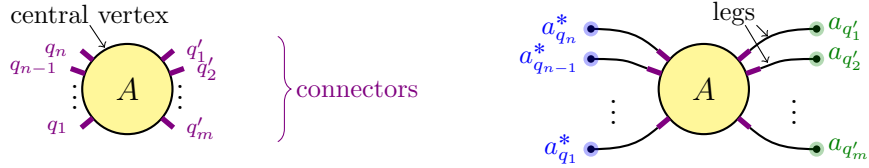
\begin{figure}[hbt]
	\centering
	\scalebox{1.0}{\def\r{0.6} 
\def\extl(#1, #2){\fill (#1, #2) circle (0.05); \fill[opacity = 0.3, blue] (#1, #2) circle (0.1);  } 
\def\extr(#1, #2){\fill (#1, #2) circle (0.05); \fill[opacity = 0.3, green!50!black] (#1, #2) circle (0.1);  } 
\def\conn(#1){({\r*cos(#1)},{\r*sin(#1)}) --  ({(\r+0.2)*cos(#1)},{(\r+0.2)*sin(#1)})} 
\def\connl(#1, #2){({\r*cos(#1)},{\r*sin(#1)}) .. controls  ({(\r+#2)*cos(#1)},{(\r+#2)*sin(#1)}) and} 
\begin{tikzpicture}
\useasboundingbox (-2,-1) rectangle (4,1.2);

\filldraw[fill = yellow!50!white, thick] (0,0) circle (\r) node{$A$} ;

\node at (-0.7,0) {$ \vdots $};
\node at (0.7,0) {$ \vdots $};

\node at (-0.5,1) {\footnotesize central vertex};
\draw[->] (-0.5,0.85) -- ++(0.2,-0.3);

\draw[line width = 2, red!50!blue] \conn(140) node[anchor = east]{\scriptsize $ q_{n} $};
\draw[line width = 2, red!50!blue] \conn(160) node[anchor = east]{\scriptsize $ q_{n-1} $};
\draw[line width = 2, red!50!blue] \conn(220) node[anchor = east]{\scriptsize $ q_{1} $};

\draw[line width = 2, red!50!blue] \conn(40) node[anchor = west]{\scriptsize $ q'_{1} $};
\draw[line width = 2, red!50!blue] \conn(20) node[anchor = west]{\scriptsize $ q'_{2} $};
\draw[line width = 2, red!50!blue] \conn(-40) node[anchor = west]{\scriptsize $ q'_{m} $};

\draw [decorate,decoration={brace, amplitude=5pt}, red!50!blue]  (2,0.7) -- ++ (0,-1.4);
\node[red!50!blue, anchor = west] at (2.1,0) {\footnotesize connectors};

\end{tikzpicture}}
	\scalebox{1.0}{\def\r{0.6} 
\def\extl(#1, #2){\fill (#1, #2) circle (0.05); \fill[opacity = 0.3, blue] (#1, #2) circle (0.1);  } 
\def\extr(#1, #2){\fill (#1, #2) circle (0.05); \fill[opacity = 0.3, green!50!black] (#1, #2) circle (0.1);  } 
\def\conn(#1){({\r*cos(#1)},{\r*sin(#1)}) --  ({(\r+0.2)*cos(#1)},{(\r+0.2)*sin(#1)})} 
\def\connl(#1, #2){({(\r+0.2)*cos(#1)},{(\r+0.2)*sin(#1)}) .. controls  ({(\r+#2)*cos(#1)},{(\r+#2)*sin(#1)}) and} 
\begin{tikzpicture}
\useasboundingbox (-2.7,-1) rectangle (2.5,1.2);

\filldraw[fill = yellow!50!white, thick] (0,0) circle (\r) node{$A$} ;

\draw[thick] \connl(140, 0.5) ++(0.5,0) .. (-1.5,0.8) node[blue, anchor = east]{\footnotesize $ a^*_{q_n} $}; \extl(-1.5, 0.8)
\draw[thick] \connl(160, 0.5) ++(0.5,0) .. (-1.5,0.4) node[blue, anchor = east]{\footnotesize $ a^*_{q_{n - 1}} $}; \extl(-1.5, 0.4)
\node at (-1.2,-0.2) {$ \vdots $};
\draw[thick] \connl(220, 0.5) ++(0.5,0) .. (-1.5,-0.8) node[blue, anchor = east]{\footnotesize $ a^*_{q_1} $}; \extl(-1.5, -0.8)

\draw[thick] \connl(40, 0.5) ++(-0.5,0) .. (1.5,0.8) node[green!50!black, anchor = west]{\footnotesize $ a_{q_1'} $} ; \extr(1.5, 0.8)
\draw[thick] \connl(20, 0.5) ++(-0.5,0) .. (1.5,0.4) node[green!50!black, anchor = west]{\footnotesize $ a_{q_2'} $}; \extr(1.5, 0.4)
\node at (1.2,-0.2) {$ \vdots $};
\draw[thick] \connl(-40, 0.5) ++(-0.5,0) .. (1.5,-0.8) node[green!50!black, anchor = west]{\footnotesize $ a_{q_m'} $}; \extr(1.5, -0.8)

\node at (0.4,1) {\footnotesize legs};
\draw[->] (0.55,0.85) -- ++(0.3,-0.5);
\draw[->] (0.7,0.85) -- ++(0.15,-0.15);

\draw[line width = 2, red!50!blue] \conn(140);
\draw[line width = 2, red!50!blue] \conn(160);
\draw[line width = 2, red!50!blue] \conn(220);

\draw[line width = 2, red!50!blue] \conn(40);
\draw[line width = 2, red!50!blue] \conn(20);
\draw[line width = 2, red!50!blue] \conn(-40);

\end{tikzpicture}}
	\caption{Left: A vertex with connectors, representing $ f $ and its momenta $ q_j, q_j' $.\\ Right: A Friedrichs diagram with one vertex, representing $ A $ in \eqref{eq:Aoperator}.}
	\label{fig:Friedrichs_A}
\end{figure}

Let us quickly recap the diagrammatic formalism by Friedrichs \cite{friedrichs1965perturbation}, for which we use the same notation as in \cite{brooks2023friedrichs}. Since we will need to evaluate both bosonic and fermionic commutators, we will assume in this section that $ a_q^*, a_q $ can describe both a species of fermionic or of bosonic creation/annihilation operators. That is, they satisfy either the CAR \eqref{eq:CAR} or the CCR
\begin{equation}
\label{eq:CCRCAR}
	[a_q, a_{q'}^*] = \delta_{q, q'} \;, \qquad
	[a_q, a_{q'}] = [a_q^*, a_{q'}^*] = 0 \qquad \text{for all } q, q' \in \ZZZ^3 \;.
\end{equation}
In the formalism of Friedrichs diagrams, an operator of the form
\begin{equation}
	A = \sum_{\substack{q_1, \ldots, q_n \\ q'_1, \ldots, q'_m}} f(q_1, \ldots, q_n, q'_1, \ldots, q'_m)
	a^*_{q_n} \ldots a^*_{q_1} a_{q'_1} \ldots a_{q'_m}
\label{eq:Aoperator}
\end{equation}
is represented by a vertex, see Fig.~\ref{fig:Friedrichs_A}, that encodes the function (``kernel'') $ f \in \ell^2(\ZZZ^{3(n+m)}) $ with $ n $ legs pointing to the left and $ m $ legs pointing to the right.

\begin{figure}[hbt]
	\centering
	\scalebox{0.9}{\def\r{0.6} 
\def\rB{0.8} 
\def\rC{0.5} 
\def\extl(#1, #2){\fill (#1, #2) circle (0.05); \fill[opacity = 0.3, blue] (#1, #2) circle (0.1);  } 
\def\extr(#1, #2){\fill (#1, #2) circle (0.05); \fill[opacity = 0.3, green!50!black] (#1, #2) circle (0.1);  } 
\def\conn(#1, #2, #3, #4){({#4*cos(#1) + #2},{#4*sin(#1) + #3}) --  ({(#4+0.2)*cos(#1) + #2},{(#4+0.2)*sin(#1) + #3})} 
\def\connt(#1, #2, #3, #4, #5){({#4*cos(#1) + #2},{#4*sin(#1) + #3}) .. controls  ({(#4+#5)*cos(#1) + #2},{(#4+#5)*sin(#1) + #3}) and} 
\def\connte(#1, #2, #3, #4, #5){ ({(#4+#5)*cos(#1) + #2},{(#4+#5)*sin(#1) + #3}) .. ({#4*cos(#1) + #2},{#4*sin(#1) + #3}) } 
\begin{tikzpicture}
\useasboundingbox (0,-2.6) rectangle (5,1.2);

\filldraw[fill = yellow!50!white, thick] (0,0) circle (\r) node{$A_1$} ;

\draw[line width = 2, red!50!blue] \conn(160, 0, 0, \r) node[anchor = east]{\scriptsize $ q_{1, 4} $};
\draw[line width = 2, red!50!blue] \conn(180, 0, 0, \r) node[anchor = east]{\scriptsize $ q_{1, 3} $};
\draw[line width = 2, red!50!blue] \conn(200, 0, 0, \r) node[anchor = east]{\scriptsize $ q_{1, 2} $};
\draw[line width = 2, red!50!blue] \conn(220, 0, 0, \r) node[anchor = east]{\scriptsize $ q_{1, 1} $};

\draw[line width = 2, red!50!blue] \conn(0, 0, 0, \r) node[anchor = west]{\scriptsize $ q'_{1, 1} $};
\draw[line width = 2, red!50!blue] \conn(-20, 0, 0, \r) node[anchor = west]{\scriptsize $ q'_{1, 2} $};
\draw[line width = 2, red!50!blue] \conn(-40, 0, 0, \r) node[anchor = west]{\scriptsize $ q'_{1, 3} $};

\filldraw[fill = yellow!50!white, thick] (3.5,-0.5) circle (\rB) node{$A_2$} ;

\draw[line width = 2, red!50!blue] \conn(140, 3.5, -0.5, \rB) node[anchor = east]{\scriptsize $ q_{2, 5} $};
\draw[line width = 2, red!50!blue] \conn(160, 3.5, -0.5, \rB) node[anchor = east]{\scriptsize $ q_{2, 4} $};
\draw[line width = 2, red!50!blue] \conn(180, 3.5, -0.5, \rB) node[anchor = east]{\scriptsize $ q_{2, 3} $};
\draw[line width = 2, red!50!blue] \conn(200, 3.5, -0.5, \rB) node[anchor = east]{\scriptsize $ q_{2, 2} $};
\draw[line width = 2, red!50!blue] \conn(220, 3.5, -0.5, \rB) node[anchor = east]{\scriptsize $ q_{2, 1} $};

\draw[line width = 2, red!50!blue] \conn(40, 3.5, -0.5, \rB) node[anchor = west]{\scriptsize $ q'_{2, 1} $};
\draw[line width = 2, red!50!blue] \conn(20, 3.5, -0.5, \rB) node[anchor = west]{\scriptsize $ q'_{2, 2} $};
\draw[line width = 2, red!50!blue] \conn(-20, 3.5, -0.5, \rB) node[anchor = west]{\scriptsize $ q'_{2, 3} $};
\draw[line width = 2, red!50!blue] \conn(-40, 3.5, -0.5, \rB) node[anchor = west]{\scriptsize $ q'_{2, 4} $};

\filldraw[fill = yellow!50!white, thick] (1,-2) circle (\rC) node{$A_3$} ;

\draw[line width = 2, red!50!blue] \conn(120, 1, -2, \rC) node[anchor = east]{\scriptsize $ q_{3, 3} $};
\draw[line width = 2, red!50!blue] \conn(150, 1, -2, \rC) node[anchor = east]{\scriptsize $ q_{3, 2} $};
\draw[line width = 2, red!50!blue] \conn(180, 1, -2, \rC) node[anchor = east]{\scriptsize $ q_{3, 1} $};

\draw[line width = 2, red!50!blue] \conn(40, 1, -2, \rC) node[anchor = west]{\scriptsize $ q'_{3, 1} $};
\draw[line width = 2, red!50!blue] \conn(0, 1, -2, \rC) node[anchor = west]{\scriptsize $ q'_{3, 2} $};

\end{tikzpicture}}
	\scalebox{0.9}{\def\r{0.6} 
\def\rB{0.8} 
\def\rC{0.5} 
\def\rG{1} 
\def\xGA{8} 
\def\xGB{12} 
\def\extl(#1, #2){\fill (#1, #2) circle (0.05); \fill[opacity = 0.3, blue] (#1, #2) circle (0.1);  } 
\def\extr(#1, #2){\fill (#1, #2) circle (0.05); \fill[opacity = 0.3, green!50!black] (#1, #2) circle (0.1);  } 
\def\conn(#1, #2, #3, #4){({#4*cos(#1) + #2},{#4*sin(#1) + #3}) --  ({(#4+0.2)*cos(#1) + #2},{(#4+0.2)*sin(#1) + #3})} 
\def\connt(#1, #2, #3, #4, #5){({#4*cos(#1) + #2},{#4*sin(#1) + #3}) .. controls  ({(#4+#5)*cos(#1) + #2},{(#4+#5)*sin(#1) + #3}) and} 
\def\connte(#1, #2, #3, #4, #5){ ({(#4+#5)*cos(#1) + #2},{(#4+#5)*sin(#1) + #3}) .. ({#4*cos(#1) + #2},{#4*sin(#1) + #3}) } 
\begin{tikzpicture}
\useasboundingbox (-2,-2.6) rectangle (4.5,1.2);

\draw[thick] \connt(140, 2.5, -0.5, (\rB+0.2), 0.8) ++(1,0) .. (0,0.8) -- (-1.5,0.8) node[blue, anchor = east]{\footnotesize $ a^*_{2, 5} $}; \extl(-1.5, 0.8)
\draw[thick] \connt(160, 0, 0, (\r+0.2), 0.3) ++(0.5,0) .. (-1.5,0.4) node[blue, anchor = east]{\footnotesize $ a^*_{1, 4} $}; \extl(-1.5, 0.4)
\draw[thick] \connt(180, 0, 0, (\r+0.2), 0.3) ++(0.5,0) .. (-1.5,0.0) node[blue, anchor = east]{\footnotesize $ a^*_{1, 3} $}; \extl(-1.5, 0)
\draw[thick] \connt(120, 1, -2, (\rC+0.2), 0.8) ++(0.5,0) .. (-1.2,-0.8) -- (-1.5,-0.8) node[blue, anchor = east]{\footnotesize $ a^*_{3, 3} $}; \extl(-1.5, -0.8)
\draw[thick] \connt(150, 1, -2, (\rC+0.2), 0.8) ++(0.5,0) .. (-1, -1.2) -- (-1.5,-1.2) node[blue, anchor = east]{\footnotesize $ a^*_{3, 2} $}; \extl(-1.5, -1.2)
\draw[thick] \connt(220, 0, 0, (\r+0.2)), 0.8) ++(1,0) .. (-1.5,-1.6) node[blue, anchor = east]{\footnotesize $ a^*_{1, 1} $}; \extl(-1.5, -1.6)
\draw[thick] \connt(180, 1, -2, (\rC+0.2), 0.3) ++(0.5,0) .. (-1.5,-2) node[blue, anchor = east]{\footnotesize $ a^*_{3, 1} $}; \extl(-1.5, -2)

\draw[thick] \connt(40, 2.5, -0.5, (\r+0.2), 0.3) ++(-0.5,0) .. (4,0.4) -- (4.3,0.4) node[green!50!black, anchor = west]{\footnotesize $ a_{2, 1} $} ; \extr(4.3, 0.4)
\draw[thick] \connt(20, 2.5, -0.5, (\r+0.2), 0.3) ++(-0.3,0) .. (4.1,0) -- (4.3,0) node[green!50!black, anchor = west]{\footnotesize $ a_{2, 2} $}; \extr(4.3, 0)
\draw[thick] \connt(0, 1, -2, (\rC+0.2), 2) ++(-0.5,0) .. (4.2,-0.8) -- (4.3,-0.8) node[green!50!black, anchor = west]{\footnotesize $ a_{3, 2} $}; \extr(4.3, -0.8)
\draw[thick] \connt(-20, 2.5, -0.5, (\rB+0.2), 0.3) ++(-0.5,0) .. (4.2,-1.2) -- (4.3,-1.2) node[green!50!black, anchor = west]{\footnotesize $ a_{2, 3} $}; \extr(4.3, -1.2)

\draw[red, opacity = .8, line width = 1] \connt(200, 0, 0, (\r+0.2), 1.8) \connte(-40, 2.5, -0.5, (\rB+0.2), 1.5);
\draw[red, opacity = .8, line width = 1] \connt(0, 0, 0, (\r+0.2), 1) \connte(220, 2.5, -0.5, (\rB+0.2), 1);
\draw[red, opacity = .8, line width = 1] \connt(-20, 0, 0, (\r+0.2), 0.3) \connte(160, 2.5, -0.5, (\rB+0.2), 0.3);
\draw[red, opacity = .8, line width = 1] \connt(-40, 0, 0, (\r+0.2), 0.3) \connte(180, 2.5, -0.5, (\rB+0.2), 0.3);
\draw[red, opacity = .8, line width = 1] \connt(40, 1, -2, (\rC+0.2), 0.3) \connte(200, 2.5, -0.5, (\rB+0.2), 0.3);

\draw[rounded corners = 30, dashed, thick, red!50!blue] (-1.2,-2.6) rectangle (4,1);

\filldraw[fill = yellow!50!white, thick] (0,0) circle (\r) node{$A_1$} ;

\draw[line width = 2, red!50!blue] \conn(160, 0, 0, \r);
\draw[line width = 2, red!50!blue] \conn(180, 0, 0, \r);
\draw[line width = 2, red!50!blue] \conn(200, 0, 0, \r);
\draw[line width = 2, red!50!blue] \conn(220, 0, 0, \r);

\draw[line width = 2, red!50!blue] \conn(0, 0, 0, \r);
\draw[line width = 2, red!50!blue] \conn(-20, 0, 0, \r);
\draw[line width = 2, red!50!blue] \conn(-40, 0, 0, \r);

\filldraw[fill = yellow!50!white, thick] (2.5,-0.5) circle (\rB) node{$A_2$} ;

\draw[line width = 2, red!50!blue] \conn(140, 2.5, -0.5, \rB);
\draw[line width = 2, red!50!blue] \conn(160, 2.5, -0.5, \rB);
\draw[line width = 2, red!50!blue] \conn(180, 2.5, -0.5, \rB);
\draw[line width = 2, red!50!blue] \conn(200, 2.5, -0.5, \rB);
\draw[line width = 2, red!50!blue] \conn(220, 2.5, -0.5, \rB);

\draw[line width = 2, red!50!blue] \conn(40, 2.5, -0.5, \rB);
\draw[line width = 2, red!50!blue] \conn(20, 2.5, -0.5, \rB);
\draw[line width = 2, red!50!blue] \conn(-20, 2.5, -0.5, \rB);
\draw[line width = 2, red!50!blue] \conn(-40, 2.5, -0.5, \rB);

\filldraw[fill = yellow!50!white, thick] (1,-2) circle (\rC) node{$A_3$} ;

\draw[line width = 2, red!50!blue] \conn(120, 1, -2, \rC);
\draw[line width = 2, red!50!blue] \conn(150, 1, -2, \rC);
\draw[line width = 2, red!50!blue] \conn(180, 1, -2, \rC);

\draw[line width = 2, red!50!blue] \conn(40, 1, -2, \rC);
\draw[line width = 2, red!50!blue] \conn(0, 1, -2, \rC);

\end{tikzpicture}}
	\caption{Left: Vertices and connectors of a Friedrichs diagram with 3 vertices.\\ Right: A Friedrichs diagram with 3 vertices.}
	\label{fig:Friedrichs_cont}
\end{figure}
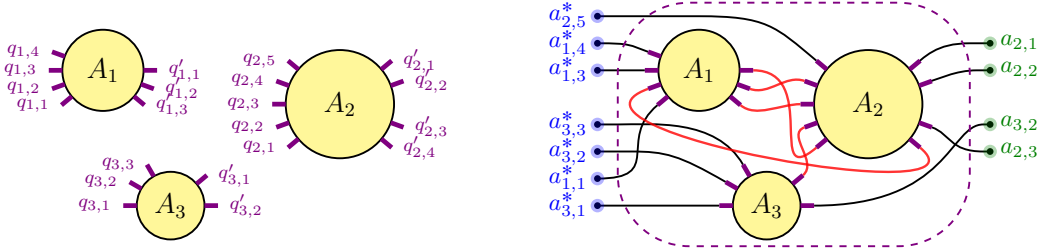

When taking multicommutators as in \eqref{eq:BCH}, the CCR/CAR will produce Kronecker deltas of the kind $ \delta_{q, q'} $, which we represent by contracted legs. To be precise, a general Friedrichs diagram as in Fig.~\ref{fig:Friedrichs_cont} consists of:
\begin{itemize}
\item[$ \bullet $] $ V $ vertices, indexed by $ v \in \{1, \ldots, V\} $, representing $ f_v: \ZZZ^{3(n_v + m_v)} \to \CCC $.
\item[$ \bullet $] $ n_v $ left-connectors and $ m_v $ right-connectors on each vertex, representing the momenta $ q_{v,1}, \ldots, q_{v,n_v} $ and $ q'_{v,1}, \ldots, q'_{v,m_v} $, respectively. So the total index sets are
\begin{equation}
\begin{aligned}
	\cJ &= \bigcup_{v = 1}^V \cJ_v, \quad
		&& \cJ_v := \{ (v, 1), \ldots, (v, n_v) \} \qquad \text{and}\\
	\cJ' &= \bigcup_{v = 1}^V \cJ'_v, \quad
		&&\cJ'_v := \{ (v, 1), \ldots, (v, m_v) \} \;.
\end{aligned}
\end{equation}
\item[$ \bullet $] $ C \le \min\{|\cJ|, |\cJ'| \} $ contractions between a left- and a right-connector. We formally keep track of them by two maps\footnote{We remark that $ \pi, \pi' $ do not play the same role as $ \pi_p, \pi_h $ above. Here, $ \pi $ indexes contracted connectors on the left and $ \pi' $ those on the right, while $ \pi_p, \pi_h $ directly associate connectors on the left to those on the right. In fact, the notation with $ \pi, \pi' $ is more general. Above, we chose the more compact notation with $ \pi_p, \pi_h $, since there we are in a special case where we know that all connectors have to be contracted.} $ \pi: \{1, \ldots, C\} \to \cJ $ and $ \pi': \{1, \ldots, C\} \to \cJ' $, where a contraction goes from connector $ \pi'(c) $ to $ \pi(c) $.
\item[$ \bullet $] $ |\cJ| + |\cJ'| - 2C $ external legs, one for each uncontracted connector, representing creation operators $ a^*_q $ (for left-connectors) or annihilation operators $ a_{q'} $ (for right-connectors). We formally keep track of the operator orderings by maps $ \sigma: \{1, \ldots, |\cJ| - C\} \to \cJ $ and $ \sigma': \{1, \ldots, |\cJ'| - C\} \to \cJ' $
\end{itemize}

\vspace{-0.5cm}

For brevity, we introduce the momentum vectors $ \bQ_v := (q_{v,1}, \ldots, q_{v,n_v}) $, $ \bQ'_v := (q'_{v,1}, \ldots, q'_{v,m_v}) $ and $ \bQ = (\bQ_1, \ldots, \bQ_V) $, $ \bQ' = (\bQ'_1, \ldots, \bQ'_V) $ and abbreviate $ a^*_{q_{v, \ell}} =: a^*_{v, \ell} $ and $ a_{q'_{v, \ell}} =: a_{v, \ell} $. A Friedrichs diagram can then be translated into an operator
\begin{equation}
	G = \sum_{\bQ, \bQ'} \left( \prod_{v = 1}^V f_v(\bQ_v, \bQ'_v) \right)
	\left( \prod_{c = 1}^C \delta_{q_{\pi(c)}, q'_{\pi'(c)}} \right)
	\left( \prod_{\ell = 1}^{|\cJ| - C} a_{\sigma(\ell)} \right)^*
	\left( \prod_{\ell' = 1}^{|\cJ'| - C} a_{\sigma'(\ell')} \right) \;.
\label{eq:G}
\end{equation}
Note that \eqref{eq:G} is again an operator of the form \eqref{eq:Aoperator}, so it can alternatively be written as a Friedrichs diagram with a single vertex.\\ 

Commutators between two bosonic or two fermionic operators $ A_1 $ and $ A_2 $ of type \eqref{eq:Aoperator} can now be expressed in terms of so-called attached products $ A_1 \cont A_2 $ (bosonic) and $ A_1 \contfer A_2 $ (fermionic): Loosely speaking, those are ``sums over all ways to contract $ A_1 $ with $ A_2 $ from left to right'', possibly including signs. Mathematically, we may track the ``ways to contract'' by the set of contraction configurations
\begin{equation}
\begin{aligned}
	\cC := \big\{ (\pi, \pi') \; \big\vert \;
	&\pi: \{1, \ldots, C \} \to \cJ_2, \;
	\pi': \{1, \ldots, C\} \to \cJ'_1, \\
	&1 \le C \le \min(m_1, n_2), \; |\mathrm{imag}(\pi')|=C,\;
	\pi(1) > \ldots > \pi(C) \big\} \;,
\end{aligned}
\label{eq:cC}
\end{equation}
where each $ (\pi, \pi') $ renders two sets of contractible but uncontracted connectors
\begin{equation}
\begin{aligned}
	\cU := &\big\{ (2, j) \in \cJ_2 \; \mid \nexists\; c \in \{1, \ldots, C\} : \pi(c) = (2, j) \big\} \;,\\
	\cU' := &\big\{ (1, k) \in \cJ'_1 \; \mid \nexists \; c \in \{1, \ldots, C\} : \pi'(c) = (1, k) \big\} \;.
\end{aligned}
\label{eq:cU}
\end{equation}
The \textbf{bosonic attached product} is then defined as
\begin{equation}
\begin{aligned}
	A_1 \cont A_2 := \sum_{(\pi, \pi') \in \cC} \sum_{\bQ, \bQ'}
	&f_1(\bQ_1, \bQ'_1) f_2(\bQ_2, \bQ'_2)
	\left( \prod_{c = 1}^C \delta_{q_{\pi(c)}, q'_{\pi'(c)}} \right) \times \\
	&\times \left(\prod_{\ell = 1}^{n_1} a_{1, \ell}\right)^*
	\left(\prod_{u \in \cU} a_u \right)^*
	\prod_{u' \in \cU'} a_{u'}
	\prod_{\ell' = 1}^{m_2} a_{2, \ell'} \;.
\end{aligned}
\label{eq:cont}
\end{equation}
For fermions, the attached product is defined analogously, up to a change of sign in front of certain contributions: Let $ \sigma, \sigma' $ be the permutations of $ \cJ_2 $ and $ \cJ'_1 $ that take the diagram into a \textbf{maximally crossed} form, while preserving the order of the uncontracted connectors. Here, by maximally crossed, we mean that the first right-connector of $ A_1 $ from the bottom, $ (1, m_1) $, is connected to the first left-connector of $ A_2 $ from the top, $ (2, n_2) $, the second to the second, and so on. That means,
\begin{equation}
\begin{aligned}
	\sigma(\pi(c)) &= (2, n_2 - c + 1) \;
	&\text{and} \quad u_1 < u_2 &\Rightarrow \sigma(u_1) < \sigma(u_2) \quad
	&&\forall \; u_1, u_2 \in \cU \;,\\
	\sigma(\pi'(c)) &= (1, m_1 - c + 1) \;
	&\text{and} \quad u'_1 < u'_2 &\Rightarrow \sigma'(u'_1) < \sigma'(u'_2) \quad
	&&\forall \; u'_1, u'_2 \in \cU' \;.
\end{aligned}
\label{eq:fullcrossing}
\end{equation}
See also Fig.~\ref{fig:Friedrichs_maxcrossing}. The sign of a contraction configuration $ (\pi, \pi') \in \cC $ is then given by
\begin{equation}
	\sgn(\pi, \pi') := (-1)^{(m_1 - C)(n_2 - C)} \sgn(\sigma) \sgn(\sigma')
\label{eq:sgnpipi}
\end{equation}
and the \textbf{fermionic attached product} is defined as
\begin{equation}
\begin{aligned}
	A_1 \contfer A_2 := \sum_{(\pi, \pi') \in \cC} \sgn(\pi, \pi') 
	&\sum_{\bQ, \bQ'} 
	f_1(\bQ_1, \bQ'_1) f_2(\bQ_2, \bQ'_2)
	\prod_{c = 1}^C \delta_{q_{\pi(c)}, q'_{\pi'(c)}} \times \\
	&\times \left( \prod_{\ell = 1}^{n_1} a_{1, \ell} \right)^*
	\left( \prod_{u \in \cU} a_u \right)^*
	\prod_{u' \in \cU'} a_{u'}
	\prod_{\ell' = 1}^{m_2} a_{2, \ell'} \;.
\end{aligned}
\label{eq:contfer}
\end{equation}
For a motivation of the sign factor $ \sgn(\pi, \pi') $, see \cite[Appendix~A]{brooks2023friedrichs}.\\

\noindent 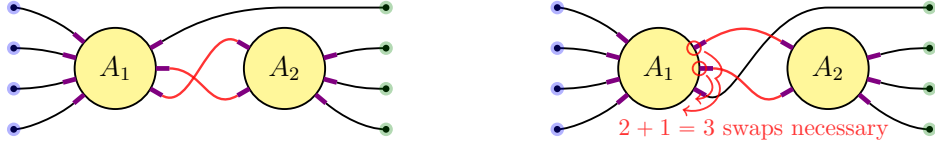
\begin{figure}
	\centering
	\hspace{-0.5cm}
	\scalebox{0.9}{\def\r{0.6} 
\def\rB{0.8} 
\def\rC{0.5} 
\def\extl(#1, #2){\fill (#1, #2) circle (0.05); \fill[opacity = 0.3, blue] (#1, #2) circle (0.1);  } 
\def\extr(#1, #2){\fill (#1, #2) circle (0.05); \fill[opacity = 0.3, green!50!black] (#1, #2) circle (0.1);  } 
\def\conn(#1, #2, #3, #4){({#4*cos(#1) + #2},{#4*sin(#1) + #3}) --  ({(#4+0.2)*cos(#1) + #2},{(#4+0.2)*sin(#1) + #3})} 
\def\connt(#1, #2, #3, #4, #5){({#4*cos(#1) + #2},{#4*sin(#1) + #3}) .. controls  ({(#4+#5)*cos(#1) + #2},{(#4+#5)*sin(#1) + #3}) and} 
\def\connte(#1, #2, #3, #4, #5){ ({(#4+#5)*cos(#1) + #2},{(#4+#5)*sin(#1) + #3}) .. ({#4*cos(#1) + #2},{#4*sin(#1) + #3}) } 
\begin{tikzpicture}
\useasboundingbox (-2,-1) rectangle (4.5,1);

\draw[thick] \connt(140, 0, 0, (\r+0.2), 0.3) ++(0.2,0) .. (-1.5,0.9); \extl(-1.5, 0.9)
\draw[thick] \connt(165, 0, 0, (\r+0.2), 0.3) ++(0.2,0) .. (-1.5,0.3); \extl(-1.5, 0.3)
\draw[thick] \connt(195, 0, 0, (\r+0.2), 0.3) ++(0.2,0) .. (-1.5,-0.3); \extl(-1.5, -0.3)
\draw[thick] \connt(220, 0, 0, (\r+0.2), 0.3) ++(0.2,0) .. (-1.5,-0.9); \extl(-1.5, -0.9)

\draw[thick] \connt(30, 0, 0, (\r+0.2), 0.3) ++(-1,0) .. (2.5,0.9) -- (4,0.9); \extr(4, 0.9)
\draw[thick] \connt(15, 2.5, 0, (\r+0.2), 0.3) ++(-0.3,0) .. (4,0.3); \extr(4, 0.3)
\draw[thick] \connt(-15, 2.5, 0, (\r+0.2), 0.3) ++(-0.3,0) .. (4,-0.3); \extr(4, -0.3)
\draw[thick] \connt(-40, 2.5, 0, (\r+0.2), 0.3) ++(-0.3,0) .. (4,-0.9); \extr(4, -0.9)

\draw[red, opacity = .8, line width = 1] \connt(0, 0, 0, (\r+0.2), 0.5) \connte(210, 2.5, 0, (\r+0.2), 0.5);
\draw[red, opacity = .8, line width = 1] \connt(-30, 0, 0, (\r+0.2), 0.5) \connte(150, 2.5, 0, (\r+0.2), 0.5);

\filldraw[fill = yellow!50!white, thick] (0,0) circle (\r) node{$A_1$} ;

\draw[line width = 2, red!50!blue] \conn(140, 0, 0, \r);
\draw[line width = 2, red!50!blue] \conn(165, 0, 0, \r);
\draw[line width = 2, red!50!blue] \conn(195, 0, 0, \r);
\draw[line width = 2, red!50!blue] \conn(220, 0, 0, \r);

\draw[line width = 2, red!50!blue] \conn(30, 0, 0, \r);
\draw[line width = 2, red!50!blue] \conn(0, 0, 0, \r);
\draw[line width = 2, red!50!blue] \conn(-30, 0, 0, \r);

\filldraw[fill = yellow!50!white, thick] (2.5,0) circle (\r) node{$A_2$} ;

\draw[line width = 2, red!50!blue] \conn(150, 2.5, 0, \r);
\draw[line width = 2, red!50!blue] \conn(210, 2.5, 0, \r);

\draw[line width = 2, red!50!blue] \conn(15, 2.5, 0, \r);
\draw[line width = 2, red!50!blue] \conn(-15, 2.5, 0, \r);
\draw[line width = 2, red!50!blue] \conn(-40, 2.5, 0, \r);

\end{tikzpicture}}
	\scalebox{0.9}{\def\r{0.6} 
\def\rB{0.8} 
\def\rC{0.5} 
\def\extl(#1, #2){\fill (#1, #2) circle (0.05); \fill[opacity = 0.3, blue] (#1, #2) circle (0.1);  } 
\def\extr(#1, #2){\fill (#1, #2) circle (0.05); \fill[opacity = 0.3, green!50!black] (#1, #2) circle (0.1);  } 
\def\conn(#1, #2, #3, #4){({#4*cos(#1) + #2},{#4*sin(#1) + #3}) --  ({(#4+0.2)*cos(#1) + #2},{(#4+0.2)*sin(#1) + #3})} 
\def\connt(#1, #2, #3, #4, #5){({#4*cos(#1) + #2},{#4*sin(#1) + #3}) .. controls  ({(#4+#5)*cos(#1) + #2},{(#4+#5)*sin(#1) + #3}) and} 
\def\connp(#1, #2, #3, #4, #5){({ #4*cos(#1) + #2},{#4*sin(#1) + #3}) } 
\def\connte(#1, #2, #3, #4, #5){ ({(#4+#5)*cos(#1) + #2},{(#4+#5)*sin(#1) + #3}) .. ({#4*cos(#1) + #2},{#4*sin(#1) + #3}) } 
\begin{tikzpicture}
\useasboundingbox (-2,-1) rectangle (4.5,1);

\draw[thick] \connt(140, 0, 0, (\r+0.2), 0.3) ++(0.2,0) .. (-1.5,0.9); \extl(-1.5, 0.9)
\draw[thick] \connt(165, 0, 0, (\r+0.2), 0.3) ++(0.2,0) .. (-1.5,0.3); \extl(-1.5, 0.3)
\draw[thick] \connt(195, 0, 0, (\r+0.2), 0.3) ++(0.2,0) .. (-1.5,-0.3); \extl(-1.5, -0.3)
\draw[thick] \connt(220, 0, 0, (\r+0.2), 0.3) ++(0.2,0) .. (-1.5,-0.9); \extl(-1.5, -0.9)

\draw[thick] \connt(-30, 0, 0, (\r+0.2), 0.5) ++(-1,0) .. (2.5,0.9) -- (4,0.9); \extr(4, 0.9)
\draw[thick] \connt(15, 2.5, 0, (\r+0.2), 0.3) ++(-0.3,0) .. (4,0.3); \extr(4, 0.3)
\draw[thick] \connt(-15, 2.5, 0, (\r+0.2), 0.3) ++(-0.3,0) .. (4,-0.3); \extr(4, -0.3)
\draw[thick] \connt(-40, 2.5, 0, (\r+0.2), 0.3) ++(-0.3,0) .. (4,-0.9); \extr(4, -0.9)

\draw[red, opacity = .8, line width = 1] \connt(30, 0, 0, (\r+0.2), 0.5) \connte(150, 2.5, 0, (\r+0.2), 0.5);
\draw[red, opacity = .8, line width = 1] \connt(0, 0, 0, (\r+0.2), 0.5) \connte(210, 2.5, 0, (\r+0.2), 0.5);

\filldraw[fill = yellow!50!white, thick] (0,0) circle (\r) node{$A_1$} ;

\draw[line width = 2, red!50!blue] \conn(140, 0, 0, \r);
\draw[line width = 2, red!50!blue] \conn(165, 0, 0, \r);
\draw[line width = 2, red!50!blue] \conn(195, 0, 0, \r);
\draw[line width = 2, red!50!blue] \conn(220, 0, 0, \r);

\draw[line width = 2, red!50!blue] \conn(30, 0, 0, \r);
\draw[line width = 2, red!50!blue] \conn(0, 0, 0, \r);
\draw[line width = 2, red!50!blue] \conn(-30, 0, 0, \r);

\filldraw[fill = yellow!50!white, thick] (2.5,0) circle (\r) node{$A_2$} ;

\draw[line width = 2, red!50!blue] \conn(150, 2.5, 0, \r);
\draw[line width = 2, red!50!blue] \conn(210, 2.5, 0, \r);

\draw[line width = 2, red!50!blue] \conn(15, 2.5, 0, \r);
\draw[line width = 2, red!50!blue] \conn(-15, 2.5, 0, \r);
\draw[line width = 2, red!50!blue] \conn(-40, 2.5, 0, \r);

\draw[red, opacity = .8, thick] \connp(0, 0, 0, \r, 0 ) circle (0.1);
\draw[red, opacity = .8, thick, ->] \connp(0, 0.1, -0.1, \r, 0 ) .. controls ++(0.2,-0.2)  and ++(0.2,0)  .. \connp(-40, 0, 0, (\r+0.1), 0) ;
\draw[red, opacity = .8, thick] \connp(30, 0, 0, \r, 0) circle (0.1);
\draw[red, opacity = .8, thick, ->] \connp(30, 0.15, -0.05, \r, 0 ) .. controls ++(0.2,-0.1)  and ++(0.2,0.1)  .. \connp(-10, 0, 0, (\r+0.25), 0) .. controls ++(0.2,-0.2)  and ++(0.4,-0.1)  .. \connp(-60, 0, 0, (\r+0.1), 0) ;
\node[red, opacity = .8] at (1.4,-0.9) {\footnotesize $2 + 1 = 3$ swaps necessary};

\end{tikzpicture}}
	\caption{Left: A maximally crossed Friedrichs diagram.\\ Right: In this diagram, a permutation $ \sigma' $ with 3 swaps is necessary to achieve a maximally crossed form. So $ \sgn(\sigma') = -1 $.}
	\label{fig:Friedrichs_maxcrossing}
\end{figure}

(Anti-)commutators can now conveniently be expressed in terms of attached products.
\begin{proposition}[{\cite[Theorems 3.1 and 3.2]{brooks2023friedrichs}}]
For bosonic operators of the form \eqref{eq:Aoperator}, we have
\begin{equation}
	[A_1, A_2] = A_1 \cont A_2 - A_2 \cont A_1 \;.
\label{eq:contthm}
\end{equation}
For fermionic operators of the form \eqref{eq:Aoperator}, we have
\begin{equation}
\begin{aligned}
	{[A_1, A_2]} &= A_1 \contfer A_2 - A_2 \contfer A_1 \qquad &&\text{\rm if} \; (m_1 n_2 + m_2 n_1) \; \text{\rm is even} \;,\\
	\{A_1, A_2\} &= A_1 \contfer A_2 + A_2 \contfer A_1 \qquad &&\text{\rm if} \; (m_1 n_2 + m_2 n_1) \; \text{\rm is odd} \;.\\
\end{aligned}
\label{eq:contferthm}
\end{equation}
\label{prop:cont}
\end{proposition}

With these commutator formulas at hand, we are ready to evaluate the multicommutator in \eqref{eq:BCH} diagrammatically.\\

\section{Multicommutator Evaluation via Friedrichs Diagrams}
\label{sec:multicomm}

Using Friedrichs diagrams, we will now evaluate the multicommutator in \eqref{eq:BCH} to derive the formula \eqref{eq:multicommutatorexact} for $ n_q $ as claimed in Theorem \ref{thm:multicommutatorexact}.\\

\begin{proof}[Proof of Theorem \ref{thm:multicommutatorexact}]
Recall \eqref{eq:BCH}:
\begin{equation*}
	n_q
	= \sum_{n = 0}^\infty \frac{1}{n!} \langle \Omega, \ad^n_S (a_q^* a_q) \Omega \rangle \;.
\end{equation*}
First, let us specify how to represent the operators $ a_q^* a_q, c_\alpha^*(k), c_\alpha(k) $, and $ S $ diagrammatically, see Fig.~\ref{fig:Friedrichs_aaccS}.

\vspace{-0.5cm}
\noindent \begin{figure}[hbt]
	\centering
	\scalebox{1.1}{\begin{tikzpicture}

\filldraw[fill = yellow!50!white, thick] (0,0.25) circle (0.15);
\draw[thick, blue] (-0.15,0.25) -- ++(-0.45,0) node[anchor = south]{\footnotesize $ q $};
\draw[thick, blue] (0.15,0.25) -- ++(0.45,0) node[anchor = south]{\footnotesize $ q $};  
\node at (0,1.6) {$ a_q^* a_q $};

\filldraw[thick, fill opacity = .1] (2.5,0.25) circle (0.15);
\draw[thick] (2.4,0.35) -- ++(-0.5,0) node[anchor = south]{\footnotesize $ p $};
\draw[thick] (2.4,0.15) -- ++(-0.5,0) node[anchor = north]{\footnotesize $ h $};
\node at (2.3,1.6) {$ c_\alpha^*(k) $};

\filldraw[thick, fill opacity = .1] (4,0.25) circle (0.15);
\draw[thick] (4.1,0.35) -- ++(0.5,0) node[anchor = south]{\footnotesize $ h $};
\draw[thick] (4.1,0.15) -- ++(0.5,0) node[anchor = north]{\footnotesize $ p $};
\node at (4.2,1.6) {$ c_\alpha(k) $};

\filldraw[thick, fill opacity = .1] (7.5,0) circle (0.15);
\filldraw[thick, fill opacity = .1] (7.5,0.5) circle (0.15);
\filldraw[thick, rounded corners = 5, fill opacity = .1] (7.25,-0.25) rectangle ++(0.5,1);
\draw[thick] (7.4,-0.1) -- ++(-0.5,0) node[anchor = east]{\scriptsize $ h' $};
\draw[thick] (7.4,0.1) -- ++(-0.5,0) node[anchor = east]{\scriptsize $ p' $};
\draw[thick] (7.4,0.4) -- ++(-0.5,0) node[anchor = east]{\scriptsize $ h $};
\draw[thick] (7.4,0.6) -- ++(-0.5,0) node[anchor = east]{\scriptsize $ p $};
\node at (8,0.5) {\footnotesize $ \alpha $};
\node at (8,0) {\footnotesize $ \alpha' $};
\node at (7.5,1.6) {$ S_+ $};

\filldraw[thick, fill opacity = .1] (9.5,0) circle (0.15);
\filldraw[thick, fill opacity = .1] (9.5,0.5) circle (0.15);
\filldraw[thick, rounded corners = 5, fill opacity = .1] (9.25,-0.25) rectangle ++(0.5,1);
\draw[thick] (9.6,-0.1) -- ++(0.5,0) node[anchor = west]{\scriptsize $ p $};
\draw[thick] (9.6,0.1) -- ++(0.5,0) node[anchor = west]{\scriptsize $ h $};
\draw[thick] (9.6,0.4) -- ++(0.5,0) node[anchor = west]{\scriptsize $ p' $};
\draw[thick] (9.6,0.6) -- ++(0.5,0) node[anchor = west]{\scriptsize $ h' $};
\node at (9,0.5) {\footnotesize $ \alpha' $};
\node at (9,0) {\footnotesize $ \alpha $};
\node at (9.5,1.6) {$ S_- $};

\end{tikzpicture}}
	\caption{From left to right: The operators $ a_q^* a_q, c_\alpha^*(k), c_\alpha(k) $ and $ S = S_+ + S_- $ are translated into Friedrichs diagrams.}
	\label{fig:Friedrichs_aaccS}
\end{figure}
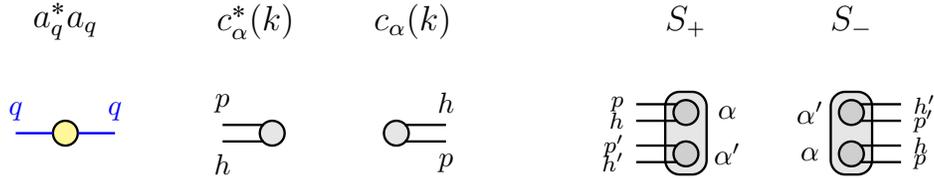

\begin{itemize}
\item[$ \bullet $] The operator $ a_q^* a_q $ is characterized by the kernel $ f(p_0, p_0') = \delta_{q,p_0} \delta_{q,p_0'} $ in case $ q \in B_{\F}^c $ and $ f(h_0, h_0') = \delta_{q,h_0} \delta_{q,h_0'} $ if $ q \in B_{\F} $. We represent it by a small vertex.

\item[$ \bullet $] Each operator $ c_\alpha^*(k) $ is characterized by a kernel
\begin{equation}
\label{eq:dalphak}
	f(p,h) = d_{\alpha, k}(p, h) := \delta_{p, h \pm k} \frac{1}{n_{\alpha, k}} \chi_{B_{\F}^c \cap B_\alpha}(p) \chi_{B_{\F} \cap B_\alpha}(h) \;,
\end{equation}
compare \eqref{eq:cstar}, where we adopted the notation of \cite[(4.2) and (4.1)]{benedikter2023momentum}. We also represent $ c_\alpha^*(k) $ by a small vertex, whose two legs are pointing left.

\item[$ \bullet $] Likewise, $ c_\alpha(k) $ is represented by a small vertex with two legs pointing to the right. The directions in which the legs are pointing makes it clear, which operator is meant by which small vertex.

\item[$ \bullet $] We represent the creation part $ S_+ $ of $ S $ by a large rectangular vertex including two smaller $ c^* $-vertices. The large vertex can then be translated into a factor of $ \frac{1}{2} K(k)_{\alpha, \alpha'} $, where sums over $ k, \alpha $, and $ \alpha' $ are implicitly assumed. Further, the large vertex fixes both momentum transfers inside the small vertices $ c^*_\alpha(k) $ and $ c^*_{\alpha'}(k) $ to be the same vector $ k \in \Gamma^{\nor} \subset \ZZZ^3 $.\\
Likewise, $ S_- $ is represented by a large vertex with 4 legs pointing to the right.\\
\end{itemize}

We would like to apply Proposition \ref{prop:cont} for evaluating the multicommutators $ \ad^n_S (a_q^* a_q) $ in \eqref{eq:BCH}. In each commutator $ \ad^n_S (a_q^* a_q) = [S_+, \ad^{n-1}_S (a_q^* a_q)] + [S_-, \ad^{n-1}_S (a_q^* a_q)] $, the leg numbers of the first vertex $ A_1 := S_\pm $ are $ (n_1, m_1) = (4,0) $ or $ (0,4) $, respectively. So irrespective of the leg numbers of the diagrams in $ A_2 := \ad^{n-1}_S (a_q^* a_q) $, we have $ m_1 n_2 + n_1 m_2 = 4 n_2 $ or $ m_1 n_2 + n_1 m_2 = 4 m_2 $, which are both even. Thus, \eqref{eq:contferthm} indeed renders a formula for a commutator.\\

Now, following Proposition \ref{prop:cont}, the multicommutators $ \ad^n_S (a_q^* a_q) $ in \eqref{eq:BCH} correspond to diagrams, which are built by starting with an $ a_q^* a_q $-vertex and successively contracting $ n $ vertices of type $ S_\pm $ into the diagram.\\
After taking the vacuum expectation value $ \langle \Omega, \ad^n_S (a_q^* a_q) \Omega \rangle $, any diagram with external legs will vanish as it yields linear combinations of terms of the form $ \langle \Omega, a_{q_n}^* \ldots a_{q_1}^* a_{q_1'} \ldots a_{q_m'} \Omega \rangle $, and we have $ a_q \Omega = 0 $. So we only need to consider diagrams where all $ 4n+2 $ legs have been contracted. As contractions always connect a left- and a right-connector, we need to have $ 2n+1 $ connectors of either kind. So only diagrams with $ \frac n2 $ vertices of type $ S_+ $ and $ \frac n2 $ vertices of type $ S_- $ contribute. In particular,
\begin{equation}
	\langle \Omega, \ad^n_S (a_q^* a_q) \Omega \rangle = 0 \qquad
	\text{if $ n $ is odd} \;,
\end{equation}
and the sum in \eqref{eq:BCH} reduces to even $ n $.\\
In order to derive \eqref{eq:multicommutatorexact}, let us back-translate the corresponding diagrams. Irrespective of the contractions, the $ S $-vertices contribute the sums $ \sum_{\bK \in (\Gamma^{\nor})^n} $ and $ \sum_{\balpha, \balpha'} $, see \eqref{eq:bKbalpha}, such that $ \alpha_j, \alpha_j' \in \cI_{k_j} $, as claimed in Theorem \ref{thm:multicommutatorexact}.\\
The $ 4n+2 $ momenta of the connectors are tracked in $ \bP, \bP', \bH, \bH' $, see \eqref{eq:bPbH}, where the condition \eqref{eq:phcondition} encodes the factors $ \chi_{B_{\F}^c \cap B_\alpha}(p) $ and $ \chi_{B_{\F} \cap B_\alpha}(h) $ from the $ c^\sharp_\alpha(k) $-vertices.\\
Now, for $ 1 \le j \le n $, every $ S_- $-vertex also contributes a factor of $ \frac 12 K(k_j)_{\alpha_j, \alpha_j'} $ and every $ S_+ $-vertex a factor of $ - \frac 12 K(k_j)_{\alpha_j, \alpha_j'} $. As $ S_+ $ contains creation operators, it gets contracted ``from the right'', yielding a contribution $ \ad^{j-1}_S (a_q^* a_q) \contfer S_+ $ which comes with an additional minus sign, see \eqref{eq:contferthm}. So including the sign from \eqref{eq:contferthm}, both $ S_+ $ and $ S_- $ effectively contribute $ \frac 12 K(k_j)_{\alpha_j, \alpha_j'} $.\\
Further, the two $ c^\sharp_\alpha(k) $-vertices in $ S_\pm $ contribute a factor of $ \delta_{p_j, h_j \pm k_j} n_{\alpha_j, k_j}^{-1} $ and $\delta_{p_j', h_j' \pm k_j} n_{\alpha_j', k_j}^{-1} $. Recalling that each order $ n \in \NNN $ comes with a factor of $ (n!)^{-1} $ in the series \eqref{eq:BCH}, this reproduces the first line of \eqref{eq:multicommutatorexact}.\\

The second line of \eqref{eq:multicommutatorexact} now accounts for contractions in the diagrams. As explained around \eqref{eq:Xin}, the map $ \xi \in \Xi_n $ tracks whether $ S_+ $ or $ S_- $ has been chosen for contraction, which is unique for each diagram.\\
Accordingly, we split the connectors into those on the right ($ \bP_+, \bP'_+, \bH_+, \bH'_+ $), and those on the left ($ \bP_-, \bP'_-, \bH_-, \bH'_- $) as in \eqref{eq:bPbHpartitions}.\\
The $ 2n+1 $ contractions are tracked by the bijective maps $ \pi_p, \pi_h $ as in \eqref{eq:pippih}. Here, $ \pi_p $ associates to every particle-connector $ p $ on the left a particle-connector $ \pi_p(p) $ on the right to which $ p $ is contracted. Every such contraction results in a contribution of $ \delta_{p, \pi_p(p)} $. Likewise, every hole-connector $ h $ on the left is contracted to $ \pi_h(h) $ on the right, resulting in a contribution of $ \delta_{h, \pi_h(h)} $.\\
As the attached product $ \contfer $ must include at least one contraction, we require the $ j $-th $ S $-vertex to be contracted to an existing $ \ell $-th vertex ($ \ell < j $). This is exactly constraint \eqref{eq:piconstraint}, resulting in the contraction sum running over $ (\pi_p, \pi_h) \in \Pi_n^{(\xi)} $.\\
Finally, the factor of $ \delta_{q,p_0} \delta_{q,p_0'} $ or $ \delta_{q,h_0} \delta_{q,h_0'} $ is just the kernel of the $ a_q^* a_q $-vertex, as explained above.\\

\vspace{-0.5cm}
\noindent 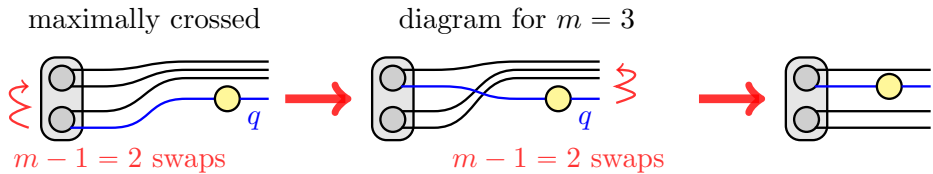
\begin{figure}[hbt]
	\centering
	\scalebox{1.1}{\begin{tikzpicture}

\node at (1.5,1.2) {\footnotesize maximally crossed};
\filldraw[thick, fill opacity = .1] (0.5,0) circle (0.15);
\filldraw[thick, fill opacity = .1] (0.5,0.5) circle (0.15);
\filldraw[thick, rounded corners = 5, fill opacity = .1] (0.25,-0.25) rectangle ++(0.5,1);
\draw[thick, blue] (0.6,-0.1) -- ++(0.5,0) .. controls ++(0.5,0) and ++(-0.5,0) .. (2,0.25) -- (2.35,0.25);
\draw[thick] (0.6,0.1) -- ++(0.4,0) .. controls ++(0.5,0) and ++(-0.5,0) .. (2,0.5) -- (3,0.5);
\draw[thick] (0.6,0.4) -- ++(0.4,0) .. controls ++(0.5,0) and ++(-0.5,0) .. (2,0.6) -- (3,0.6);
\draw[thick] (0.6,0.6) -- ++(0.4,0) .. controls ++(0.5,0) and ++(-0.5,0) .. (2,0.7) -- (3,0.7);

\filldraw[fill = yellow!50!white, thick] (2.5,0.25) circle (0.15);
\draw[thick, blue] (2.65,0.25) -- node[anchor = north]{\footnotesize $ q $} ++(0.35,0);  

\draw[red, opacity = .8, thick, ->] (0.1,-0.1) .. controls ++(-0.3,0.1) and ++(-0.3,-0.1) .. (0.1,0.15) .. controls ++(-0.3,0.1) and ++(-0.3,0) .. (0.1,0.4);
\draw[red, opacity = .8, line width = 3, ->] (3.2,0.25) -- ++(0.8,0);
\node[red, opacity = .8] at (1.2,-0.5) {\footnotesize $ m-1 = 2 $ swaps};

\node at (6,1.2) {\footnotesize diagram for $ m=3 $};
\filldraw[thick, fill opacity = .1] (4.5,0) circle (0.15);
\filldraw[thick, fill opacity = .1] (4.5,0.5) circle (0.15);
\filldraw[thick, rounded corners = 5, fill opacity = .1] (4.25,-0.25) rectangle ++(0.5,1);
\draw[thick] (4.6,-0.1) -- ++(0.4,0) .. controls ++(0.5,0) and ++(-0.5,0) .. (6,0.5) -- (7,0.5);
\draw[thick] (4.6,0.1) -- ++(0.4,0) .. controls ++(0.5,0) and ++(-0.5,0) .. (6,0.6) -- (7,0.6);
\draw[thick, blue] (4.6,0.4) -- ++(0.5,0) .. controls ++(0.5,0) and ++(-0.5,0) .. (6,0.25) -- (6.35,0.25);
\draw[thick] (4.6,0.6) -- ++(0.4,0) .. controls ++(0.5,0) and ++(-0.5,0) .. (6,0.7) -- (7,0.7);

\filldraw[fill = yellow!50!white, thick] (6.5,0.25) circle (0.15);
\draw[thick, blue] (6.65,0.25) -- node[anchor = north]{\footnotesize $ q $} ++(0.35,0);  

\draw[red, opacity = .8, thick, ->] (7.2,0.2) .. controls ++(0.3,0.1) and ++(0.3,-0.1) .. (7.2,0.4) .. controls ++(0.3,0.1) and ++(0.3,0) .. (7.2,0.6);
\node[red, opacity = .8] at (6.5,-0.5) {\footnotesize $ m-1 = 2 $ swaps};

\draw[red, opacity = .8, line width = 3, ->] (8.2,0.25) -- ++(0.8,0);

\filldraw[thick, fill opacity = .1] (9.5,0) circle (0.15);
\filldraw[thick, fill opacity = .1] (9.5,0.5) circle (0.15);
\filldraw[thick, rounded corners = 5, fill opacity = .1] (9.25,-0.25) rectangle ++(0.5,1);
\draw[thick] (9.6,-0.1) -- ++(1.4,0);
\draw[thick] (9.6,0.1) -- ++(1.4,0);
\draw[thick, blue] (9.6,0.4) -- ++(0.75,0);
\draw[thick] (9.6,0.6) -- ++(1.4,0);

\filldraw[fill = yellow!50!white, thick] (10.5,0.4) circle (0.15);
\draw[thick, blue] (10.65,0.4) -- ++(0.35,0); 

\end{tikzpicture}} 
	\caption{After the first contraction, the diagram can be brought into the same structure as $ S_\pm $, while picking up a sign factor of $ (-1)^{2m-2} = 1 $.}
	\label{fig:Firstcontraction}
\end{figure}

The only remaining step is to evaluate the sign factors $ \sgn(\pi, \pi') $ as in \eqref{eq:sgnpipi} and to show that their product amounts to the factor $ \sgn(\xi, \pi_p, \pi_h) $ in \eqref{eq:sgnxipippih}. We start with considering the first contraction, which appears in $ \ad^1_S (a_q^* a_q) = [S_+, a_q^* a_q] + [S_-, a_q^* a_q] $. Here, one $ q $-connector gets contracted to any of the 4 connectors of $ S_\pm $, say, the $ m $-th one, counted from the top for $ S_+ $ and from the bottom for $ S_- $. The diagram then arises from a maximally crossed one by employing $ m-1 $ swaps, see Fig.~\ref{fig:Firstcontraction}. If we now swap the remaining uncontracted $ a_q^\sharp $-operator into the position, where the contracted operator of $ S_\pm $ used to be (second step in Fig.~\ref{fig:Firstcontraction}), we have to apply further $ m-1 $ swaps. After this move, we end up with an operator that has the same structure as an $ S_\pm $-operator, while the sign factor we pick up is $ (-1)^{2m-2} = 1 $. So for the sign evaluation, we can proceed as if the $ a_q^* a_q $-vertex didn't exist.\\

\vspace{-0.5cm}
\noindent \begin{figure}[hbt]
	\centering
	\scalebox{1.0}{\begin{tikzpicture}

\filldraw[fill = yellow!50!white, thick] (-1.0,-0.1) circle (0.15);

\filldraw[thick, blue, fill opacity = .1] (-1.5,0) circle (0.15);
\filldraw[thick, fill opacity = .1] (-1.5,0.5) circle (0.15);
\filldraw[thick, rounded corners = 5, fill opacity = .1] (-1.75,-0.25) rectangle ++(0.5,1);
\draw[thick, blue] (-1.4,-0.1) -- ++(0.25,0);
\draw[thick, blue] (-0.85,-0.1) -- ++(0.15,0) node[anchor = west] {\tiny $ q $};
\draw[thick] (-1.4,0.1) -- ++(0.7,0) node[anchor = west] {\tiny $ h_1 $};
\draw[thick] (-1.4,0.4) -- ++(0.7,0) node[anchor = west] {\tiny $ p_1' $};
\draw[thick] (-1.4,0.6) -- ++(0.7,0) node[anchor = west] {\tiny $ h_1' $};

\filldraw[thick, fill opacity = .1] (-1.5,1.25) circle (0.15);
\filldraw[thick, fill opacity = .1] (-1.5,1.75) circle (0.15);
\filldraw[thick, rounded corners = 5, fill opacity = .1] (-1.75,1) rectangle ++(0.5,1);
\draw[thick] (-1.4,1.15) -- ++(0.7,0) node[anchor = west] {\tiny $ p_3 $};
\draw[thick] (-1.4,1.35) -- ++(0.7,0) node[anchor = west] {\tiny $ h_3 $};
\draw[thick] (-1.4,1.65) -- ++(0.7,0) node[anchor = west] {\tiny $ p_3' $};
\draw[thick] (-1.4,1.85) -- ++(0.7,0) node[anchor = west] {\tiny $ h_3' $};

\filldraw[thick, fill opacity = .1] (-1.5,2.5) circle (0.15);
\filldraw[thick,  fill opacity = .1] (-1.5,3) circle (0.15);
\filldraw[thick, rounded corners = 5, fill opacity = .1] (-1.75,2.25) rectangle ++(0.5,1);
\draw[thick] (-1.4,2.4) -- ++(0.7,0) node[anchor = west] {\tiny $ p_6 $};
\draw[thick] (-1.4,2.6) -- ++(0.7,0) node[anchor = west] {\tiny $ h_6 $};
\draw[thick] (-1.4,2.9) -- ++(0.7,0) node[anchor = west] {\tiny $ p_6' $};
\draw[thick] (-1.4,3.1) -- ++(0.7,0) node[anchor = west] {\tiny $ h_6' $};

\filldraw[thick, fill opacity = .1] (1.5,0) circle (0.15);
\filldraw[thick, fill opacity = .1] (1.5,0.5) circle (0.15);
\filldraw[thick, rounded corners = 5, fill opacity = .1] (1.25,-0.25) rectangle ++(0.5,1);
\draw[thick] (1.4,-0.1) -- ++(-0.7,0) node[anchor = east] {\tiny $ h_5' $};
\draw[thick] (1.4,0.1) -- ++(-0.7,0) node[anchor = east] {\tiny $ p_5' $};
\draw[thick] (1.4,0.4) -- ++(-0.7,0) node[anchor = east] {\tiny $ h_5 $};
\draw[thick] (1.4,0.6) -- ++(-0.7,0) node[anchor = east] {\tiny $ p_5 $};

\filldraw[thick, fill opacity = .1] (1.5,1.25) circle (0.15);
\filldraw[thick, fill opacity = .1] (1.5,1.75) circle (0.15);
\filldraw[thick, rounded corners = 5, fill opacity = .1] (1.25,1) rectangle ++(0.5,1);
\draw[thick] (1.4,1.15) -- ++(-0.7,0) node[anchor = east] {\tiny $ h_4' $};
\draw[thick] (1.4,1.35) -- ++(-0.7,0) node[anchor = east] {\tiny $ p_4' $};
\draw[thick] (1.4,1.65) -- ++(-0.7,0) node[anchor = east] {\tiny $ h_4 $};
\draw[thick] (1.4,1.85) -- ++(-0.7,0) node[anchor = east] {\tiny $ p_4 $};

\filldraw[thick, fill opacity = .1] (1.5,2.5) circle (0.15);
\filldraw[thick, fill opacity = .1] (1.5,3) circle (0.15);
\filldraw[thick, rounded corners = 5, fill opacity = .1] (1.25,2.25) rectangle ++(0.5,1);
\draw[thick] (1.4,2.4) -- ++(-0.7,0) node[anchor = east] {\tiny $ h_2' $};
\draw[thick] (1.4,2.6) -- ++(-0.7,0) node[anchor = east] {\tiny $ p_2' $};
\draw[thick] (1.4,2.9) -- ++(-0.7,0) node[anchor = east] {\tiny $ h_2 $};
\draw[thick] (1.4,3.1) -- ++(-0.7,0) node[anchor = east] {\tiny $ p_2 $};

\node at (-2,0) {\footnotesize $ \alpha_1 $};
\node at (-2,0.5) {\footnotesize $ \alpha_1' $};
\node at (-2,1.25) {\footnotesize $ \alpha_3 $};
\node at (-2,1.75) {\footnotesize $ \alpha_3' $};
\node at (-2,2.5) {\footnotesize $ \alpha_6 $};
\node at (-2,3) {\footnotesize $ \alpha_6' $};

\node at (2,0) {\footnotesize $ \alpha_5' $};
\node at (2,0.5) {\footnotesize $ \alpha_5 $};
\node at (2,1.25) {\footnotesize $ \alpha_4' $};
\node at (2,1.75) {\footnotesize $ \alpha_4 $};
\node at (2,2.5) {\footnotesize $ \alpha_2' $};
\node at (2,3) {\footnotesize $ \alpha_2 $};

\draw[->] (-2.4,0.5) -- ++(0,2);
\draw[->] (2.4,2.5) -- ++(0,-2);

\end{tikzpicture}} \hspace{0.8cm}
	\scalebox{1.0}{\begin{tikzpicture}

\filldraw[fill = yellow!50!white, thick] (-0.5,-0.1) circle (0.15);

\filldraw[thick, blue, fill opacity = .1] (-1,0) circle (0.15);
\filldraw[thick, fill opacity = .1] (-1,0.5) circle (0.15);
\filldraw[thick, rounded corners = 5, fill opacity = .1] (-1.25,-0.25) rectangle ++(0.5,1);
\draw[thick, blue] (-0.9,-0.1) -- ++(0.25,0);
\draw[thick, blue] (-0.35,-0.1) -- ++(0.15,0);
\draw[thick, blue] (-0.9,0.1) -- ++(0.3,0);
\draw[thick] (-0.9,0.4) -- ++(0.3,0);
\draw[thick] (-0.9,0.6) -- ++(0.3,0);

\filldraw[thick, fill opacity = .1] (-1,1.25) circle (0.15);
\filldraw[thick, fill opacity = .1] (-1,1.75) circle (0.15);
\filldraw[thick, rounded corners = 5, fill opacity = .1] (-1.25,1) rectangle ++(0.5,1);
\draw[thick] (-0.9,1.15) -- ++(0.3,0);
\draw[thick] (-0.9,1.35) -- ++(0.3,0) .. controls ++(0.5,0) and ++(-0.5,0) .. (0.5,1.15) -- ++(0.4,0);
\draw[thick] (-0.9,1.65) -- ++(0.3,0);
\draw[thick] (-0.9,1.85) -- ++(0.3,0);

\filldraw[thick, fill opacity = .1] (-1,2.5) circle (0.15);
\filldraw[thick,  fill opacity = .1] (-1,3) circle (0.15);
\filldraw[thick, rounded corners = 5, fill opacity = .1] (-1.25,2.25) rectangle ++(0.5,1);
\draw[thick] (-0.9,2.4) -- ++(0.3,0);
\draw[thick] (-0.9,2.6) -- ++(0.3,0);
\draw[thick] (-0.9,2.9) -- ++(0.3,0) .. controls ++(0.5,0) and ++(-0.5,0) .. (0.4,1.85) -- ++(0.5,0);
\draw[thick] (-0.9,3.1) -- ++(0.3,0);

\filldraw[thick, fill opacity = .1] (1,0) circle (0.15);
\filldraw[thick, fill opacity = .1] (1,0.5) circle (0.15);
\filldraw[thick, rounded corners = 5, fill opacity = .1] (0.75,-0.25) rectangle ++(0.5,1);
\draw[thick] (0.9,-0.1) -- ++(-0.3,0);
\draw[thick] (0.9,0.1) -- ++(-0.3,0);
\draw[thick] (0.9,0.4) -- ++(-0.3,0);
\draw[thick] (0.9,0.6) -- ++(-0.3,0);

\filldraw[thick, fill opacity = .1] (1,1.25) circle (0.15);
\filldraw[thick, fill opacity = .1] (1,1.75) circle (0.15);
\filldraw[thick, rounded corners = 5, fill opacity = .1] (0.75,1) rectangle ++(0.5,1);
\draw[thick] (0.9,1.35) -- ++(-0.3,0);
\draw[thick] (0.9,1.65) -- ++(-0.3,0);

\filldraw[thick, fill opacity = .1] (1,2.5) circle (0.15);
\filldraw[thick, fill opacity = .1] (1,3) circle (0.15);
\filldraw[thick, rounded corners = 5, fill opacity = .1] (0.75,2.25) rectangle ++(0.5,1);
\draw[thick] (0.9,2.4) -- ++(-0.3,0);
\draw[thick] (0.9,2.6) -- ++(-0.3,0);
\draw[thick] (0.9,2.9) -- ++(-0.3,0);
\draw[thick] (0.9,3.1) -- ++(-0.3,0);

\node at (-1.7,1.35) {\footnotesize $ q_- $};
\node at (-1.7,2.9) {\footnotesize $ q_-' $};
\node at (1.9,1.15) {\footnotesize $ \pi_{\sharp}(q_-) $};
\node at (1.9,1.85) {\footnotesize $ \pi_{\sharp}(q_-') $};

\end{tikzpicture}} \hspace{0.8cm}
	\scalebox{1.0}{\begin{tikzpicture}

\filldraw[fill = yellow!50!white, thick] (-0.5,-0.1) circle (0.15);

\filldraw[thick, blue, fill opacity = .1] (-1,0) circle (0.15);
\filldraw[thick, fill opacity = .1] (-1,0.5) circle (0.15);
\filldraw[thick, rounded corners = 5, fill opacity = .1] (-1.25,-0.25) rectangle ++(0.5,1);
\draw[thick, blue] (-0.9,-0.1) -- ++(0.25,0);
\draw[thick, blue] (-0.35,-0.1) -- ++(0.15,0) .. controls ++(0.5,0) and ++(-0.5,0) .. (0.7,3.1) -- ++(0.2,0);
\draw[thick, blue] (-0.9,0.1) -- ++(0.7,0) .. controls ++(0.5,0) and ++(-0.5,0) .. (0.7,2.9) -- ++(0.2,0);
\draw[thick] (-0.9,0.4) -- ++(0.7,0) .. controls ++(0.5,0) and ++(-0.5,0) .. (0.7,2.6) -- ++(0.2,0);
\draw[thick] (-0.9,0.6) -- ++(0.7,0) .. controls ++(0.5,0) and ++(-0.5,0) .. (0.7,2.4) -- ++(0.2,0);

\filldraw[thick, fill opacity = .1] (-1,1.25) circle (0.15);
\filldraw[thick, fill opacity = .1] (-1,1.75) circle (0.15);
\filldraw[thick, rounded corners = 5, fill opacity = .1] (-1.25,1) rectangle ++(0.5,1);
\draw[thick] (-0.9,1.15) -- ++(0.5,0) .. controls ++(0.5,0) and ++(-0.5,0) .. (0.5,1.85) -- ++(0.4,0);
\draw[thick] (-0.9,1.35) -- ++(0.5,0) .. controls ++(0.5,0) and ++(-0.5,0) .. (0.5,1.65) -- ++(0.4,0);
\draw[thick] (-0.9,1.65) -- ++(0.5,0) .. controls ++(0.5,0) and ++(-0.5,0) .. (0.5,1.35) -- ++(0.4,0);
\draw[thick] (-0.9,1.85) -- ++(0.5,0) .. controls ++(0.5,0) and ++(-0.5,0) .. (0.5,1.15) -- ++(0.4,0);

\filldraw[thick, fill opacity = .1] (-1,2.5) circle (0.15);
\filldraw[thick,  fill opacity = .1] (-1,3) circle (0.15);
\filldraw[thick, rounded corners = 5, fill opacity = .1] (-1.25,2.25) rectangle ++(0.5,1);
\draw[thick] (-0.9,2.4) -- ++(0.2,0) .. controls ++(0.5,0) and ++(-0.5,0) .. (0.5,0.6) -- ++(0.4,0);
\draw[thick] (-0.9,2.6) -- ++(0.2,0) .. controls ++(0.5,0) and ++(-0.5,0) .. (0.5,0.4) -- ++(0.4,0);
\draw[thick] (-0.9,2.9) -- ++(0.2,0) .. controls ++(0.5,0) and ++(-0.5,0) .. (0.5,0.1) -- ++(0.4,0);
\draw[thick] (-0.9,3.1) -- ++(0.2,0) .. controls ++(0.5,0) and ++(-0.5,0) .. (0.5,-0.1) -- ++(0.4,0);

\filldraw[thick, fill opacity = .1] (1,0) circle (0.15);
\filldraw[thick, fill opacity = .1] (1,0.5) circle (0.15);
\filldraw[thick, rounded corners = 5, fill opacity = .1] (0.75,-0.25) rectangle ++(0.5,1);

\filldraw[thick, fill opacity = .1] (1,1.25) circle (0.15);
\filldraw[thick, fill opacity = .1] (1,1.75) circle (0.15);
\filldraw[thick, rounded corners = 5, fill opacity = .1] (0.75,1) rectangle ++(0.5,1);

\filldraw[thick, fill opacity = .1] (1,2.5) circle (0.15);
\filldraw[thick, fill opacity = .1] (1,3) circle (0.15);
\filldraw[thick, rounded corners = 5, fill opacity = .1] (0.75,2.25) rectangle ++(0.5,1);

\end{tikzpicture}} 
	\caption{Left: Example of a diagram with $ n=6 $ vertices with only the first contraction drawn.\\
	Middle: A situation with $ q_-' > q_- $ and $ \pi_{\sharp}(q_-') < \pi_{\sharp}(q_-) $, making a swap necessary to finally achieve maximal crossing.\\
	Right: A maximally crossed diagram with $ n=6 $.}
	\label{fig:Friedrichs_multicommutator}
\end{figure}
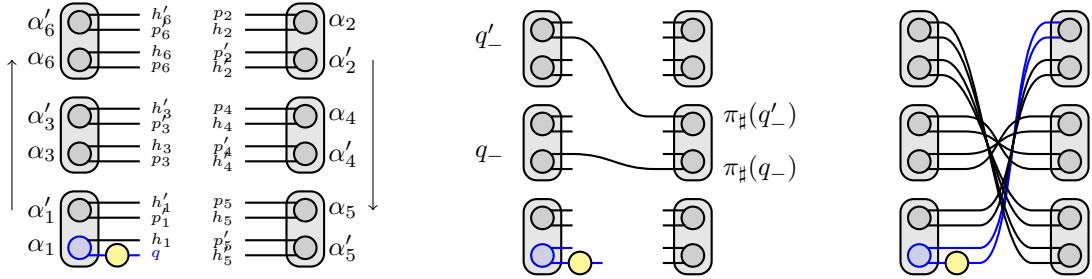

In the following contraction steps, we then successively add $ S_- $-diagrams from the bottom to the top on the left-hand side and $ S_+ $-diagrams from the top to the bottom on the right-hand side, see Fig.~\ref{fig:Friedrichs_multicommutator}. Note that the order (bottom to top or vice versa) is enforced by the ordering prescription introduced in \eqref{eq:Aoperator} and below. Also, observe that the ordering relation $ q_-' > q_- $ in \eqref{eq:orderingrelation} and \eqref{eq:sgnxipippih} just means that the connector $ q_-' $ is above $ q_- $ in the diagram, while $ \pi_{\sharp}(q_-') < \pi_{\sharp}(q_-) $ means that the connector $ \pi_{\sharp}(q_-') $ is above $ \pi_{\sharp}(q_-) $, see also Fig.~\ref{fig:Friedrichs_multicommutator}. So the sign factor $ \sgn(\xi, \pi_p, \pi_h) $ in \eqref{eq:sgnxipippih} essentially counts how many swaps would be necessary to take the diagram into maximally crossed form\footnote{Note that the maximally crossed diagram in Fig.~\ref{fig:Friedrichs_multicommutator} does not contribute to $ n_q $ due to the constraint \eqref{eq:piconstraint}.} as depicted in Fig.~\ref{fig:Friedrichs_multicommutator}, while ignoring the $ a_q^* a_q $-vertex. Thus, we may finish the proof by establishing the following claim.\\

\noindent \underline{Claim}: The product of all sign factors $ \sgn(\pi, \pi') $ appearing in the multicommutator evaluation for every diagram indexed by $ (\xi, \pi_p, \pi_h) $ is identical to the sign factor $ \sgn(\xi, \pi_p, \pi_h) $ we would need to make the diagram, including $ n $ vertices $ S_\pm $ and ignoring $ a_q^* a_q $, maximally crossed.\\

\noindent \underline{Proof of the Claim}: Following \eqref{eq:sgnpipi}, a factor of $ (-1) $ in $ \sgn(\pi, \pi') $ enters if and only if within the contraction of a new $ S_\pm $-vertex:
\begin{itemize}
\item[$ (A) $] A connector gets contracted, and it has to be swapped with another connector in order to achieve maximal crossing. This rule accounts for the factor of $ \sgn(\sigma) \sgn(\sigma') $ in \eqref{eq:sgnpipi}.
\item[$ (B) $] A connector of some operator $ a $ is not contracted and has to ``jump'' over another uncontracted operator $ a^* $ to achieve normal ordering. This accounts for $ (-1)^{(m_1 - C)(n_2 - C)} $ in \eqref{eq:sgnpipi}.
\end{itemize}

\vspace{-0.5cm}
\noindent 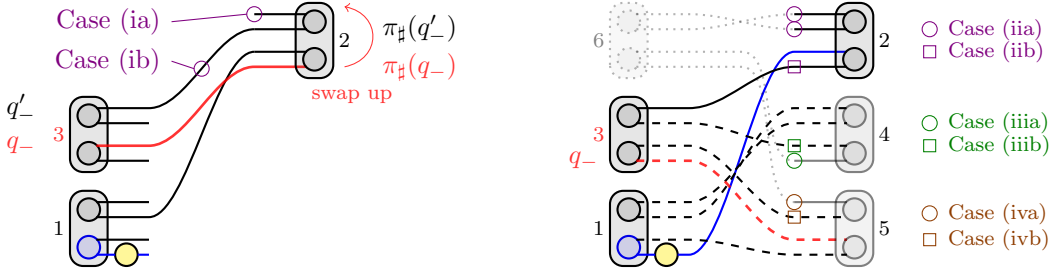
\begin{figure}[hbt]
	\centering
	\scalebox{1.0}{\begin{tikzpicture}

\filldraw[fill = yellow!50!white, thick] (-1.0,-0.1) circle (0.15);

\filldraw[thick, blue, fill opacity = .1] (-1.5,0) circle (0.15);
\filldraw[thick, fill opacity = .1] (-1.5,0.5) circle (0.15);
\filldraw[thick, rounded corners = 5, fill opacity = .1] (-1.75,-0.25) rectangle ++(0.5,1);
\draw[thick, blue] (-1.4,-0.1) -- ++(0.25,0);
\draw[thick, blue] (-0.85,-0.1) -- ++(0.15,0);
\draw[thick] (-1.4,0.1) -- ++(0.7,0);
\draw[thick] (-1.4,0.4) -- ++(0.7,0) .. controls ++(0.5,0) and ++(-0.5,0) .. (0.7,2.6);
\draw[thick] (-1.4,0.6) -- ++(0.7,0);

\filldraw[thick, fill opacity = .1] (-1.5,1.25) circle (0.15);
\filldraw[thick, fill opacity = .1] (-1.5,1.75) circle (0.15);
\filldraw[thick, rounded corners = 5, fill opacity = .1] (-1.75,1) rectangle ++(0.5,1);
\draw[thick] (-1.4,1.15) -- ++(0.7,0);
\draw[red, opacity = 0.8, line width = 1] (-1.4,1.35) -- ++(0.7,0) .. controls ++(0.5,0) and ++(-0.5,0) .. (0.7,2.4) -- ++(0.7,0);
\draw[thick] (-1.4,1.65) -- ++(0.7,0);
\draw[thick] (-1.4,1.85) -- ++(0.7,0) .. controls ++(0.5,0) and ++(-0.5,0) .. (0.7,2.9);

\filldraw[thick, fill opacity = .1] (1.5,2.5) circle (0.15);
\filldraw[thick, fill opacity = .1] (1.5,3) circle (0.15);
\filldraw[thick, rounded corners = 5, fill opacity = .1] (1.25,2.25) rectangle ++(0.5,1);
\draw[thick] (1.4,2.6) -- ++(-0.7,0);
\draw[thick] (1.4,2.9) -- ++(-0.7,0);
\draw[thick] (1.4,3.1) -- ++(-0.7,0);

\node[red, opacity = 0.8] at (-2.4,1.35) {\footnotesize $ q_- $};
\node[red, opacity = 0.8] at (2.9,2.4) {\footnotesize $ \pi_{\sharp} (q_-) $};
\node at (-2.4,1.85) {\footnotesize $ q_-' $};
\node at (2.9,2.9) {\footnotesize $ \pi_{\sharp} (q_-') $};
\draw[red, opacity = 0.8, ->] (1.9,2.4) .. controls ++(0.5,0.1) and ++(0.5,-0.1) .. (1.9,3.2);
\node[red, opacity = 0.8, ->] at (2,2) {\scriptsize swap up};

\draw[blue!50!red] (0.7,3.1) circle (0.1);
\draw[blue!50!red] (0.6,3.1) -- ++(-1.0,-0.1) node[anchor = east] {\footnotesize Case (ia)};
\draw[blue!50!red] (0,2.375) circle (0.1);
\draw[blue!50!red] (-0.1,2.375) -- ++(-0.3,0.1) node[anchor = east] {\footnotesize Case (ib)};

\node at (-1.9,0.25) {\scriptsize 1};
\node[red, opacity = .8] at (-1.9,1.5) {\scriptsize 3};
\node at (1.9,2.75) {\scriptsize 2};

\end{tikzpicture}} \hspace{0.8cm}
	\scalebox{1.0}{\begin{tikzpicture}

\filldraw[fill = yellow!50!white, thick] (-1.0,-0.1) circle (0.15);

\filldraw[thick, blue, fill opacity = .1] (-1.5,0) circle (0.15);
\filldraw[thick, fill opacity = .1] (-1.5,0.5) circle (0.15);
\filldraw[thick, rounded corners = 5, fill opacity = .1] (-1.75,-0.25) rectangle ++(0.5,1);
\draw[thick, blue] (-1.4,-0.1) -- ++(0.25,0);
\draw[thick, blue] (-0.85,-0.1) -- ++(0.15,0) .. controls ++(0.5,0) and ++(-0.5,0) .. (0.7,2.6) -- ++(0.7,0);
\draw[dashed, thick] (-1.4,0.1) -- ++(0.7,0) .. controls ++(0.5,0) and ++(-0.5,0) .. (0.7,-0.1) -- ++(0.7,0);
\draw[dashed, thick] (-1.4,0.4) -- ++(0.7,0) .. controls ++(0.5,0) and ++(-0.5,0) .. (0.7,1.85) -- ++(0.7,0);
\draw[dashed, thick] (-1.4,0.6) -- ++(0.7,0) .. controls ++(0.5,0) and ++(-0.5,0) .. (0.7,1.65) -- ++(0.7,0);

\filldraw[thick, fill opacity = .1] (-1.5,1.25) circle (0.15);
\filldraw[thick, fill opacity = .1] (-1.5,1.75) circle (0.15);
\filldraw[thick, rounded corners = 5, fill opacity = .1] (-1.75,1) rectangle ++(0.5,1);
\draw[red, opacity = .8, dashed, line width = 1] (-1.4,1.15) -- ++(0.7,0) .. controls ++(0.5,0) and ++(-0.5,0) .. (0.7,0.1) -- ++(0.7,0);
\draw[dashed, thick] (-1.4,1.35) -- ++(0.7,0) .. controls ++(0.5,0) and ++(-0.5,0) .. (0.7,0.4) -- ++(0.7,0);
\draw[dashed, thick] (-1.4,1.65) -- ++(0.7,0) .. controls ++(0.5,0) and ++(-0.5,0) .. (0.7,1.35) -- ++(0.7,0);
\draw[thick] (-1.4,1.85) -- ++(0.7,0) .. controls ++(0.5,0) and ++(-0.5,0) .. (0.7,2.4) -- ++(0.7,0);

\filldraw[opacity = .3, dotted, thick, fill opacity = .03] (-1.5,2.5) circle (0.15);
\filldraw[opacity = .3, dotted, thick,  fill opacity = .03] (-1.5,3) circle (0.15);
\filldraw[opacity = .3, dotted, thick, rounded corners = 5, fill opacity = .03] (-1.75,2.25) rectangle ++(0.5,1);
\draw[opacity = .3, dotted, thick] (-1.4,2.4) -- ++(0.7,0) -- ++(0.7,0) .. controls ++(0.5,0) and ++(-0.5,0) .. (0.7,0.6);
\draw[opacity = .3, dotted, thick] (-1.4,2.6) -- ++(0.7,0) -- ++(0.7,0) .. controls ++(0.5,0) and ++(-0.5,0) .. (0.7,1.15);
\draw[opacity = .3, dotted, thick] (-1.4,2.9) -- ++(0.7,0) .. controls ++(0.5,0) and ++(-0.5,0) .. (0.7,3.1);
\draw[opacity = .3, dotted, thick] (-1.4,3.1) -- ++(0.7,0) .. controls ++(0.5,0) and ++(-0.5,0) .. (0.7,2.9);

\filldraw[opacity = .5, thick, fill opacity = .05] (1.5,0) circle (0.15);
\filldraw[opacity = .5, thick, fill opacity = .05] (1.5,0.5) circle (0.15);
\filldraw[opacity = .5, thick, rounded corners = 5, fill opacity = .05] (1.25,-0.25) rectangle ++(0.5,1);
\draw[opacity = .5, thick] (1.4,0.6) -- ++(-0.7,0);

\filldraw[opacity = .5, thick, fill opacity = .07] (1.5,1.25) circle (0.15);
\filldraw[opacity = .5, thick, fill opacity = .07] (1.5,1.75) circle (0.15);
\filldraw[opacity = .5, thick, rounded corners = 5, fill opacity = .07] (1.25,1) rectangle ++(0.5,1);
\draw[opacity = .5, thick] (1.4,1.15) -- ++(-0.7,0);

\filldraw[thick, fill opacity = .1] (1.5,2.5) circle (0.15);
\filldraw[thick, fill opacity = .1] (1.5,3) circle (0.15);
\filldraw[thick, rounded corners = 5, fill opacity = .1] (1.25,2.25) rectangle ++(0.5,1);
\draw[thick] (1.4,2.9) -- ++(-0.7,0);
\draw[thick] (1.4,3.1) -- ++(-0.7,0);

\node[red, opacity = .8] at (-2.1,1.15) {\footnotesize $ q_- $};

\draw[red!50!blue] (0.7,3.1) circle (0.1);
\draw[red!50!blue] (0.7,2.9) circle (0.1);
\draw[red!50!blue] (0.62,2.32) rectangle ++(0.16,0.16);

\draw[green!50!black] (0.7,1.15) circle (0.1);
\draw[green!50!black] (0.62,1.27) rectangle ++(0.16,0.16);

\draw[green!30!red!80!black] (0.7,0.6) circle (0.1);
\draw[green!30!red!80!black] (0.62,0.32) rectangle ++(0.16,0.16);

\draw[red!50!blue] (2.5,2.9) circle (0.1);
\node[red!50!blue, anchor = west] at (2.6,2.9) {\scriptsize Case (iia)};
\draw[red!50!blue] (2.42,2.52) rectangle ++(0.16,0.16);
\node[red!50!blue, anchor = west] at (2.6,2.6) {\scriptsize Case (iib)};

\draw[green!50!black] (2.5,1.65) circle (0.1);
\node[green!50!black, anchor = west] at (2.6,1.65) {\scriptsize Case (iiia)};
\draw[green!50!black] (2.42,1.27) rectangle ++(0.16,0.16);
\node[green!50!black, anchor = west] at (2.6,1.35) {\scriptsize Case (iiib)};

\draw[green!30!red!80!black] (2.5,0.45) circle (0.1);
\node[green!30!red!80!black, anchor = west] at (2.6,0.45) {\scriptsize Case (iva)};
\draw[green!30!red!80!black] (2.42,0.02) rectangle ++(0.16,0.16);
\node[green!30!red!80!black, anchor = west] at (2.6,0.1) {\scriptsize Case (ivb)};

\node at (-1.9,0.25) {\scriptsize 1};
\node[red, opacity = .8] at (-1.9,1.5) {\scriptsize 3};
\node[opacity = .5] at (-1.9,2.75) {\scriptsize 6};

\node at (1.9,0.25) {\scriptsize 5};
\node at (1.9,1.5) {\scriptsize 4};
\node at (1.9,2.75) {\scriptsize 2};

\end{tikzpicture}}
	\caption{Left: Cases (ia) and (ib), in which $ q_- $ is accounted for a factor of $ (-1) $.\\
	Right: Cases (ii), (iii) and (iv). The numbers 1--6 indicate the order in which $ S_\pm $-vertices enter the diagram. Here, $ j=3 $.}
	\label{fig:Friedrichs_itoiv}
\end{figure}

We now re-distribute the sign factors $ (-1) $ coming from (A) and (B) among the $ 2n $ connectors on the left-hand side, $ q_- \in \{p_j, p'_j, h_j, h'_j\} \subset \bQ_- $, by making every $ q_- $ accountable for the following factors of $ (-1) $, see Fig.~\ref{fig:Friedrichs_itoiv}:

\begin{itemize}
\item[(i)] If $ q_- $ gets contracted to $ \pi_{\sharp}(q_-) $ immediately as the $ j $-th vertex $ S_\pm = S_- $ enters, we make it accountable for all $ (-1) $ of type $ (A) $ caused by $ \pi_{\sharp}(q_-) $ being swapped upwards past all connectors above $ \pi_{\sharp}(q_-) $ that are 
\begin{itemize}
\item[(ia)] uncontracted or
\item[(ib)] contracted to some connector $ q_-' $ above $ q_- $.
\end{itemize}
\item[(ii)] If $ q_- $ does not immediately get contracted as the $ j $-th vertex $ S_\pm = S_- $ enters, we make it accountable for 
\begin{itemize}
\item[(iia)] all $ (-1) $ of type $ (B) $ caused by $ q_- $ jumping over uncontracted connectors on the right and
\item[(iib)] all $ (-1) $ of type $ (A) $ caused by connectors $ q_-' > q_- $ being swapped down below $ q_- $.
\end{itemize}
\item[(iii)] As an $ \ell $-th vertex $ S_\pm = S_+, \ell > j $ enters, and $ q_- $ is still not contracted, we make it accountable for
\begin{itemize}
\item[(iiia)] all $ (-1) $ of type $ (B) $ caused by $ q_- $ jumping over uncontracted connectors in $ S_- $ and
\item[(iiib)] all $ (-1) $ of type $ (A) $ caused by connectors $ q_-' > q_- $ that got contracted to vertex $ \ell $ and have to be swapped down past $ q_- $.
\end{itemize}
\item[(iv)] As an $ \ell $-th vertex $ S_\pm = S_+, \ell > j $ enters, and $ q_- $ is contracted to $ \pi_{\sharp}(q_-) $ within this step, we make $ q_- $ accountable for $ \pi_{\sharp}(q_-) $ being swapped past
\begin{itemize}
\item[(iva)] all uncontracted connectors in the $ \ell $-th $ S_\pm $ above $ \pi_{\sharp}(q_-) $
\item[(ivb)] all connectors in the $ \ell $-th $ S_\pm $ above $ \pi_{\sharp}(q_-) $ contracted to some $ q_-' > q_- $.
\end{itemize}

\end{itemize}

Indeed, multiplying all sign factors $ (-1) $ from (i)--(iv) for all $ q_- $ recovers exactly contributions (A) and (B): It is easy to see that, as the $ j $-th $ S $-vertex enters, and it is of type $ S_- $, then the contributions (iia) from all uncontracted $ q_- $ in this $ S_- $ make up all sign factors of type $ (B) $ occurring in this step. Further, (ia) and (ib) account for all swaps on the right and (iib) for all swaps on the left that are needed to achieve maximal crossing in this step, rendering all sign factors of $ (A) $.\\
By contrast, if an $ S_+ $-vertex enters, then (iiia) will yield all sign factors of type $ (B) $ in this step. Then, (iiib) accounts for all swaps on the left, and (iva) and (ivb) for all swaps on the right that yield maximal crossing after the step, rendering the factors $ (A) $. So all factors (i)--(iv) from all $ q_- $ indeed render the total product of all sign factors $ \sgn(\pi, \pi') $ appearing in the $ n $ contractions.\\

Now, let us determine the sign factor $ \sgn(\xi, \pi_p, \pi_h) $ appearing when swapping the entire diagram into maximally crossed form, as in Fig.~\ref{fig:Friedrichs_multicommutator}. The transition to maximal crossing can be achieved by successively considering connectors $ q_- \in \bQ_- $ and swapping $ \pi_{\sharp}(q_-) $ on the right past all $ \pi_{\sharp}(q_-') < \pi_{\sharp}(q_-) $ with $ q_-' > q_- $.\\
First, assume that $ q_- $ is contracted immediately as the $ j $-th vertex $ S_\pm $ enters. Any $ q_-' > q_- $ with $ \pi_{\sharp}(q_-') < \pi_{\sharp}(q_-) $ is either in the $ j $-th vertex, rendering (ib), or above the $ j $-th vertex, so $ \pi_{\sharp}(q_-') $ is not yet contracted in step $ j $, rendering (ia).\\
Conversely, assume that $ q_- $ is not immediately contracted as the $ j $-th vertex $ S_\pm $ enters, but only in step $ \ell > j $. Then, for the connectors $ \pi_{\sharp}(q_-') $ swapped with $ \pi_{\sharp}(q_-) $ there are the following options: $ \pi_{\sharp}(q_-') $ could be present in the diagram after step $ j $ and at this point either already contracted to some $ q_-' > q_- $, rendering (iib), or uncontracted, so it gets later contracted to $ q_-' $ in some $ S_- $-vertex above $ q_- $, rendering contribution (iia). Or, $ \pi_{\sharp}(q_-') $ could join the diagram in steps $ j+1 $ through $ \ell $, rendering contributions (iiib) and (ivb) if it is immediately contracted to some $ q_-' > q_- $, or contributions (iiia) and (iva) if it is yet uncontracted upon entering and gets later contracted to a $ q_-' $ in some vertex above the $ j $-th. After step $ \ell $, only connectors $ \pi_{\sharp}(q_-') $ below $ \pi_{\sharp}(q_-) $ enter, which do not contribute to the swaps encoded in \eqref{eq:sgnxipippih}.\\
Concluding both cases, we observe that multiplying all sign factors (i)--(iv) for all $ q_- $ also renders exactly the factor needed to swap the entire diagram into maximally crossed form, viz. $ \sgn(\xi, \pi_p, \pi_h) $. This establishes the claim and concludes the proof.\\

\end{proof}

\section{Bosonization Approximation in Friedrichs Diagrams}
\label{sec:bosonization}

Formula \eqref{eq:multicommutatorexact} is rather bulky, as it contains many contributions of various forms. Therefore, it makes sense to restrict to those Friedrichs diagrams, which we expect to make the largest contribution. These are exactly the diagrams corresponding to the bosonization approximation in \cite{benedikter2023momentum}, as we will see in this section.

\subsection{Heuristic Motivation}
\label{subsec:heuristics}

\begin{figure}[hbt]
	\centering
	\hspace{-0.2cm}
	\scalebox{0.9}{\begin{tikzpicture}

\node at (0,3.6) {3 loops};

\filldraw[fill = yellow!50!white, thick] (-0.5,-0.1) circle (0.15);

\filldraw[thick, blue, fill opacity = .1] (-1.5,0) circle (0.15);
\filldraw[thick, blue, fill opacity = .1] (-1.5,0.5) circle (0.15);
\filldraw[thick, rounded corners = 5, fill opacity = .1] (-1.75,-0.25) rectangle ++(0.5,1);
\filldraw[thick, blue!50!red, fill opacity = .1] (-1.5,1.25) circle (0.15);
\filldraw[thick, blue!50!red, fill opacity = .1] (-1.5,1.75) circle (0.15);
\filldraw[thick, rounded corners = 5, fill opacity = .1] (-1.75,1) rectangle ++(0.5,1);
\filldraw[thick, fill opacity = .1] (-1.5,2.5) circle (0.15);
\filldraw[thick, blue, fill opacity = .1] (-1.5,3) circle (0.15);
\filldraw[thick, rounded corners = 5, fill opacity = .1] (-1.75,2.25) rectangle ++(0.5,1);

\filldraw[thick, blue, fill opacity = .1] (1.5,0) circle (0.15);
\filldraw[thick, blue, fill opacity = .1] (1.5,0.5) circle (0.15);
\filldraw[thick, rounded corners = 5, fill opacity = .1] (1.25,-0.25) rectangle ++(0.5,1);
\filldraw[thick, fill opacity = .1] (1.5,1.25) circle (0.15);
\filldraw[thick, blue!50!red, fill opacity = .1] (1.5,1.75) circle (0.15);
\filldraw[thick, rounded corners = 5, fill opacity = .1] (1.25,1) rectangle ++(0.5,1);
\filldraw[thick, blue, fill opacity = .1] (1.5,2.5) circle (0.15);
\filldraw[thick, blue!50!red, fill opacity = .1] (1.5,3) circle (0.15);
\filldraw[thick, rounded corners = 5, fill opacity = .1] (1.25,2.25) rectangle ++(0.5,1);

\draw[thick, blue] (-1.4,-0.1) -- (-0.65,-0.1);
\draw[thick, blue] (-1.4,0.1) .. controls ++(2,0) and ++(-1,0) .. (1.4,-0.1);
\draw[thick, blue] (1.4,0.1) .. controls ++(-1.5,0) and ++(1.5,0) .. (-1.4,0.4);
\draw[thick, blue] (-1.4,0.6) .. controls ++(1.5,0) and ++(-1.5,0) .. (1.4,2.6);
\draw[thick, blue] (1.4,2.4) .. controls ++(-1.5,0) and ++(1.5,0) .. (-1.4,3.1);
\draw[thick, blue] (-1.4,2.9) .. controls ++(1.5,0) and ++(-1.5,0) .. (1.4,0.6);
\draw[thick, blue] (1.4,0.4) .. controls ++(-1,0) and ++(1,0) .. (-0.35,-0.1);
\node[blue] at (-2,0) {\footnotesize $ \alpha_q $};
\node[blue] at (2,0) {\footnotesize $ \alpha_q $};
\node[blue] at (-2,0.5) {\footnotesize $ \alpha_q $};
\node[blue] at (2,2.5) {\footnotesize $ \alpha_q $};
\node[blue] at (-2,3) {\footnotesize $ \alpha_q $};
\node[blue] at (2,0.5) {\footnotesize $ \alpha_q $};

\draw[thick, blue!50!red] (-1.4,1.35) .. controls ++(1.5,0) and ++(-1.5,0) .. (1.4,1.65);
\draw[thick, blue!50!red] (1.4,1.85) .. controls ++(-1.5,0) and ++(1.5,0) .. (-1.4,1.65);
\draw[thick, blue!50!red] (-1.4,1.85) .. controls ++(1.5,0) and ++(-1.5,0) .. (1.4,2.9);
\draw[thick, blue!50!red] (1.4,3.1) .. controls ++(-1.5,0) and ++(1.5,0) .. (-1.4,1.15);
\node[blue!50!red] at (-2,1.25) {\footnotesize $ \alpha_1 $};
\node[blue!50!red] at (2,1.75) {\footnotesize $ \alpha_1 $};
\node[blue!50!red] at (-2,1.75) {\footnotesize $ \alpha_1 $};
\node[blue!50!red] at (2,3) {\footnotesize $ \alpha_1 $};

\draw[thick] (-1.4,2.6) .. controls ++(1.5,0) and ++(-1.5,0) .. (1.4,1.15);
\draw[thick] (1.4,1.35) .. controls ++(-1.5,0) and ++(1.5,0) .. (-1.4,2.4);
\node at (-2,2.5) {\footnotesize $ \alpha_2 $};
\node at (2,1.25) {\footnotesize $ \alpha_2 $};

\end{tikzpicture}} \hspace{0.3cm}
	\scalebox{0.9}{\begin{tikzpicture}

\node at (0,3.6) {6 loops};

\filldraw[fill = yellow!50!white, thick] (-0.5,-0.1) circle (0.15);

\filldraw[thick, blue, fill opacity = .1] (-1.5,0) circle (0.15);
\filldraw[thick, blue!50!red, fill opacity = .1] (-1.5,0.5) circle (0.15);
\filldraw[thick, rounded corners = 5, fill opacity = .1] (-1.75,-0.25) rectangle ++(0.5,1);
\filldraw[thick, green, fill opacity = .1] (-1.5,1.25) circle (0.15);
\filldraw[thick, green!70!black, fill opacity = .1] (-1.5,1.75) circle (0.15);
\filldraw[thick, rounded corners = 5, fill opacity = .1] (-1.75,1) rectangle ++(0.5,1);
\filldraw[thick, fill opacity = .1] (-1.5,2.5) circle (0.15);
\filldraw[thick, green!40!black, fill opacity = .1] (-1.5,3) circle (0.15);
\filldraw[thick, rounded corners = 5, fill opacity = .1] (-1.75,2.25) rectangle ++(0.5,1);

\filldraw[thick, blue, fill opacity = .1] (1.5,0) circle (0.15);
\filldraw[thick, green, fill opacity = .1] (1.5,0.5) circle (0.15);
\filldraw[thick, rounded corners = 5, fill opacity = .1] (1.25,-0.25) rectangle ++(0.5,1);
\filldraw[thick, fill opacity = .1] (1.5,1.25) circle (0.15);
\filldraw[thick, blue!50!red, fill opacity = .1] (1.5,1.75) circle (0.15);
\filldraw[thick, rounded corners = 5, fill opacity = .1] (1.25,1) rectangle ++(0.5,1);
\filldraw[thick, green!40!black, fill opacity = .1] (1.5,2.5) circle (0.15);
\filldraw[thick, green!70!black, fill opacity = .1] (1.5,3) circle (0.15);
\filldraw[thick, rounded corners = 5, fill opacity = .1] (1.25,2.25) rectangle ++(0.5,1);

\draw[thick, blue] (-1.4,-0.1) -- (-0.65,-0.1);
\draw[thick, blue] (-1.4,0.1) .. controls ++(2,0) and ++(-1,0) .. (1.4,-0.1);
\draw[thick, blue] (1.4,0.1) .. controls ++(-1,0) and ++(1,0) .. (-0.35,-0.1);
\node[blue] at (-2,0) {\footnotesize $ \alpha_q $};
\node[blue] at (2,0) {\footnotesize $ \alpha_q $};

\draw[thick, blue!50!red] (-1.4,0.6) .. controls ++(1.5,0) and ++(-1.5,0) .. (1.4,1.65);
\draw[thick, blue!50!red] (1.4,1.85) .. controls ++(-1.5,0) and ++(1.5,0) .. (-1.4,0.4);
\node[blue!50!red] at (-2,0.5) {\footnotesize $ \alpha_1 $};
\node[blue!50!red] at (2,1.75) {\footnotesize $ \alpha_1 $};

\draw[thick] (-1.4,2.6) .. controls ++(1.5,0) and ++(-1.5,0) .. (1.4,1.15);
\draw[thick] (1.4,1.35) .. controls ++(-1.5,0) and ++(1.5,0) .. (-1.4,2.4);
\node at (-2,2.5) {\footnotesize $ \alpha_2 $};
\node at (2,1.25) {\footnotesize $ \alpha_2 $};

\draw[thick, green!40!black] (-1.4,3.1) .. controls ++(1.5,0) and ++(-1.5,0) .. (1.4,2.4);
\draw[thick, green!40!black] (1.4,2.6) .. controls ++(-1.5,0) and ++(1.5,0) .. (-1.4,2.9);
\node[green!40!black] at (-2,3) {\footnotesize $ \alpha_3 $};
\node[green!40!black] at (2,2.5) {\footnotesize $ \alpha_3 $};

\draw[thick, green!70!black] (-1.4,1.85) .. controls ++(1.5,0) and ++(-1.5,0) .. (1.4,2.9);
\draw[thick, green!70!black] (1.4,3.1) .. controls ++(-1.5,0) and ++(1.5,0) .. (-1.4,1.65);
\node[green!70!black] at (-2,1.75) {\footnotesize $ \alpha_4 $};
\node[green!70!black] at (2,3) {\footnotesize $ \alpha_4 $};

\draw[thick, green] (-1.4,1.35) .. controls ++(1.5,0) and ++(-1.5,0) .. (1.4,0.4);
\draw[thick, green] (1.4,0.6) .. controls ++(-1.5,0) and ++(1.5,0) .. (-1.4,1.15);
\node[green] at (-2,1.25) {\footnotesize $ \alpha_5 $};
\node[green] at (2,0.5) {\footnotesize $ \alpha_5 $};

\end{tikzpicture}} \hspace{0.3cm}
	\scalebox{0.9}{\begin{tikzpicture}

\filldraw[fill = yellow!50!white, thick] (-0.5,-0.1) circle (0.15);

\filldraw[thick, blue, fill opacity = .1] (-1.5,0) circle (0.15);
\filldraw[thick, blue!50!red, fill opacity = .1] (-1.5,0.5) circle (0.15);
\filldraw[thick, rounded corners = 5, fill opacity = .1] (-1.75,-0.25) rectangle ++(0.5,1);
\filldraw[thick, green, fill opacity = .1] (-1.5,1.25) circle (0.15);
\filldraw[thick, green!70!black, fill opacity = .1] (-1.5,1.75) circle (0.15);
\filldraw[thick, rounded corners = 5, fill opacity = .1] (-1.75,1) rectangle ++(0.5,1);
\filldraw[thick, fill opacity = .1] (-1.5,2.5) circle (0.15);
\filldraw[thick, green!40!black, fill opacity = .1] (-1.5,3) circle (0.15);
\filldraw[thick, rounded corners = 5, fill opacity = .1] (-1.75,2.25) rectangle ++(0.5,1);

\filldraw[thick, blue, fill opacity = .1] (1.5,0) circle (0.15);
\filldraw[thick, green, fill opacity = .1] (1.5,0.5) circle (0.15);
\filldraw[thick, rounded corners = 5, fill opacity = .1] (1.25,-0.25) rectangle ++(0.5,1);
\filldraw[thick, fill opacity = .1] (1.5,1.25) circle (0.15);
\filldraw[thick, blue!50!red, fill opacity = .1] (1.5,1.75) circle (0.15);
\filldraw[thick, rounded corners = 5, fill opacity = .1] (1.25,1) rectangle ++(0.5,1);
\filldraw[thick, green!40!black, fill opacity = .1] (1.5,2.5) circle (0.15);
\filldraw[thick, green!70!black, fill opacity = .1] (1.5,3) circle (0.15);
\filldraw[thick, rounded corners = 5, fill opacity = .1] (1.25,2.25) rectangle ++(0.5,1);

\draw[thick, blue] (-1.4,-0.1) -- (-0.65,-0.1);
\draw[thick, blue] (-1.4,0.1) .. controls ++(2,0) and ++(-1,0) .. (1.4,-0.1);
\draw[thick, blue] (1.4,0.1) .. controls ++(-1,0) and ++(1,0) .. (-0.35,-0.1);
\node[blue] at (-2,0) {\footnotesize $ \alpha_q $};
\node[blue] at (2,0) {\footnotesize $ \alpha_q $};

\draw[line width = 1, dashed, blue!50!red] (-1.35,0.5) .. controls ++(1.5,0) and ++(-1.5,0) .. (1.35,1.75);
\node[blue!50!red] at (-2,0.5) {\footnotesize $ \alpha_1 $};
\node[blue!50!red] at (2,1.75) {\footnotesize $ \alpha_1 $};

\draw[line width = 1, dashed] (-1.35,2.5) .. controls ++(1.5,0) and ++(-1.5,0) .. (1.35,1.25);
\node at (-2,2.5) {\footnotesize $ \alpha_2 $};
\node at (2,1.25) {\footnotesize $ \alpha_2 $};

\draw[line width = 1, dashed, green!40!black] (-1.35,3) .. controls ++(1.5,0) and ++(-1.5,0) .. (1.35,2.5);
\node[green!40!black] at (-2,3) {\footnotesize $ \alpha_3 $};
\node[green!40!black] at (2,2.5) {\footnotesize $ \alpha_3 $};

\draw[line width = 1, dashed, green!70!black] (-1.35,1.75) .. controls ++(1.5,0) and ++(-1.5,0) .. (1.35,3);
\node[green!70!black] at (-2,1.75) {\footnotesize $ \alpha_4 $};
\node[green!70!black] at (2,3) {\footnotesize $ \alpha_4 $};

\draw[line width = 1, dashed, green] (-1.35,1.25) .. controls ++(1.5,0) and ++(-1.5,0) .. (1.35,0.5);
\node[green] at (-2,1.25) {\footnotesize $ \alpha_5 $};
\node[green] at (2,0.5) {\footnotesize $ \alpha_5 $};

\end{tikzpicture}}
	\caption{Left: A generic Friedrichs diagram with 3 loops of lengths 6, 4, and 2.\\ Middle: A Friedrichs diagram with maximal number of loops 6, all of length 2.\\ Right: Bosonization---fermionic contraction pairs have been replaced by bosonic contractions.}
	\label{fig:Friedrichs_loop}
\end{figure}
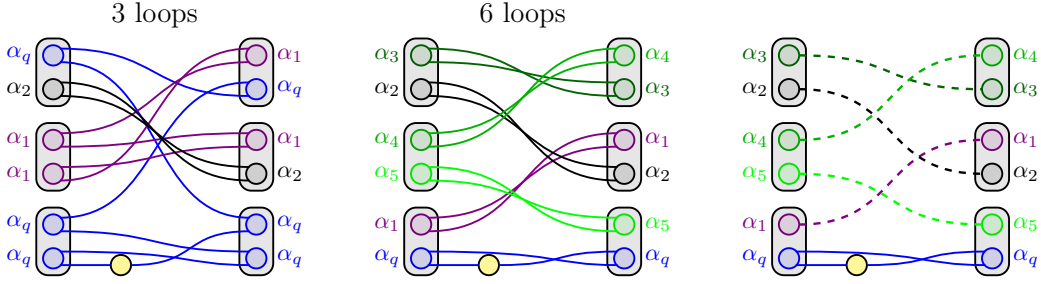

To get a heuristic intuition of which diagrams make the largest contribution, consider the first two Friedrichs diagrams in Fig.~\ref{fig:Friedrichs_loop}, which both contribute to $ \langle \Omega, \ad^6_S (a_q^* a_q) \Omega \rangle $. As a contraction $ \delta_{q_-, \pi_{\sharp}(q_-)} $ sets the momenta of the two adjacent connectors $ q_- $ and $ \pi_{\sharp}(q_-) $ equal, it also sets the patch index of the corresponding $ c^\sharp $-vertices equal to some $ \alpha_j $. The set of $ c^\sharp $-vertices thus decays into subsets of size $ \ge 2 $ with identical patch indices. In each subset of $ m $ $ c^\sharp $-vertices, the $ m $ adjacent contractions form a single loop that successively runs through all vertices. For instance, the first diagram in Fig.~\ref{fig:Friedrichs_loop} contains 3 loops, which run through $ m = 6 $, 4 and 2 $ c^\sharp $-vertices, respectively. The loop with 6 $ c^\sharp $-vertices also runs through $ a_q^* a_q $, so its patch index is fixed to $ \alpha_q $. The two shorter loops carry patch indices $ \alpha_1 $ and $ \alpha_2 $, over which we have to take a double sum $ \sum_{\alpha_1, \alpha_2} $.\\

As each sum $ \sum_{\alpha_j} $ contains $ \sim M $ terms, we expect the largest contributions to come from diagrams with the largest loop number. This is achieved if all loops\footnote{The length of a loop refers to the number of $ c^\sharp $-vertices it runs through, not taking into consideration $ a_q^* a_q $.} have length 2, resulting in $ n $ loops, as depicted in the second diagram in Fig.~\ref{fig:Friedrichs_loop}. The diagrammatic contribution then contains an $ (n-1) $-fold sum $ \sum_{\alpha_1, \ldots , \alpha_{n-1}} $. In that case, the two contractions in each loop (except the one running through the $ a_q^* a_q $-vertex) form a pair and effectively act like a single bosonic contraction, as depicted in the third diagram in Fig.~\ref{fig:Friedrichs_loop}.\\
We can thus think of the restriction to diagrams where only loops of length 2 are permitted as some kind of bosonization. These diagrams turn out to be particularly easy to evaluate and are expected to give the largest contribution to $ n_q $.\\

\noindent 

\subsection{Evaluating the Bosonized Multicommutator Diagrammatically}
\label{subsec:bosonizationevaluation}

\begin{proof}[Proof of Proposition \ref{prop:multicommutatorbos}]

We directly evaluate the right-hand side of \eqref{eq:multicommutatorbos} diagrammatically and show that it amounts to the $ \cosh $-term in \eqref{eq:nqb}. First, notice that the constraint \eqref{eq:piconstraintstrict}, in the language of the previous subsection, exactly means that we restrict to bosonized diagrams with only loops including 2 $ c^\sharp $-vertices. Thus, the right-hand side of \eqref{eq:multicommutatorbos} is just a sum over all bosonized diagrams.\\
Let us evaluate these diagrams (compare also Fig.~\ref{fig:Friedrichs_loop}). First, note that there are $ n $ loops, where in each loop, the contractions set the adjacent patch indices equal. The patch index of the loop involving the $ a_q^* a_q $-vertex is fixed to $ \alpha_q $. So the patch index sum $ \sum_{\balpha, \balpha'} $ reduces to a sum over $ n-1 $ loops, namely those in $ \balpha \setminus \alpha_j $. The contractions in a loop also set all momentum transfers $ k_j $ of the adjacent $ S_\pm $-vertices equal. Now observe that all bosonized diagrams are fully connected, since otherwise constraint \eqref{eq:piconstraint} is violated. Thus, all $ k_j $ are set equal to one single momentum transfer $ k \in \Gamma^{\nor} $ and the $ n $ -fold sum $ \sum_{\bK} $ reduces to a single sum $ \sum_k $. The contributing $ K $-matrix elements then become $ K(k)_{\alpha_j, \beta(\alpha_j)} $ with $ \beta: \balpha \to \balpha $ being an appropriate cyclic permutation.\\
Further, by the bosonization assumption \eqref{eq:piconstraintstrict}, $ p_j^\sharp = \pi_p(p_j^\sharp) $ already implies $ h_j^\sharp = \pi_h(h_j^\sharp) $ so the product $ \prod_{h \in \bH_- \cup \bH_-'} \delta_{h, \pi_h(h)} $ in \eqref{eq:multicommutatorbos} becomes redundant and can be eliminated. Also, the factor $ \delta_{p_j, h_j \pm k_j} \delta_{p_j', h_j' \pm k_j} = \delta_{p_j, h_j \pm k} \delta_{p_j', h_j' \pm k} $ in \eqref{eq:multicommutatorbos} eliminates the sums over $ \bH, \bH' $, while leaving the condition that $ h_j = p_j \mp k $ and $ h_j' = p_j' \mp k $ be holes in patch $ \alpha_j $. Let us denote these conditions as
\begin{equation}
\label{chibHbH}
	\chi(\bH, \bH' : \balpha)
	:= \prod_{j = 1}^n \chi(h_j, h_j' : \alpha_j)
	= \prod_{j = 1}^n \chi(p_j \mp k \in B_{\alpha_j} \cap B_{\F})
	\chi(p_j' \mp k \in B_{\alpha_j} \cap B_{\F}) \;.
\end{equation}
So the r.h.s. of \eqref{eq:multicommutatorbos} becomes
\begin{equation}
\label{eq:rhs1}
\begin{aligned}
	\mathrm{r.h.s.}
	= &\sum_{\substack{n = 2 \\ n : \mathrm{even}}}^\infty \frac{1}{2^n n!}
	\sum_{\xi \in \Xi_n} \sum_{(\pi_p, \pi_h) \in \Pi_{n, (\bos)}^{(\xi)}}
	\sum_k \sum_{\balpha \setminus \alpha_q}
	\sum_{\bP, \bP'}
	\chi(\bH, \bH' : \balpha)
	\times\\
	&\times
	 \left( \prod_{p \in \bP_- \cup \bP'_-} \delta_{p, \pi_p(p)} \right)
	 \left( \prod_{j = 1}^n \frac{1}{n_{\alpha_j, k}^2} K(k)_{\alpha_j, \beta(\alpha_j)} \right)
	\delta_{q, p_0} \delta_{q, p_0'} \sgn(\xi, \pi_p, \pi_h) \;.
\end{aligned}
\end{equation}
Now, the contractions $ \delta_{p, \pi_p(p)} $, not involving $ p_0 $, eliminate the sums in $ \bP, \bP' $ over all connectors on the right, that is, over all momentum indices $ p_j \in \bP_+ \setminus p_0 $ and $ p_j' \in \bP'_+ $, or equivalently, all $ p_j, p_j' $ with $ \xi(j) = 1 $. So only those $ p_j, p_j' $ on the left ($ \xi(j) = -1 $) and $ p_0, p_0' $ survive. The condition $ h_j, h_j' : \alpha_j $ is then automatically fulfilled for all $ j $ with $ \xi(j) = 1 $, so we only need to impose it on those $ j $ with $ \xi(j) = -1 $:
\begin{equation}
\begin{aligned}
	&\sum_{\bP, \bP'} \chi(\bH, \bH' : \balpha)
		\left( \prod_{p \in \bP_- \cup \bP'_-} \delta_{p, \pi_p(p)} \right)
		\delta_{q, p_0} \delta_{q, p_0'}\\
	= &\sum_{p_0, p_0'}
		\left( \prod_{j : \xi(j) = -1}
		\sum_{p_j, p_j'} \chi(h_j, h_j' : \alpha_j) \right)
		\delta_{\pi_p^{-1}(p_0), p_0}
		\delta_{q, p_0} \delta_{q, p_0'} \;.
\end{aligned}
\end{equation}
The sums $ \sum_{p_j, p_j'} $ run over all particle--hole pairs and thus amount to $ n_{\alpha_j, k}^2 $, except for the sum over $ \pi_p^{-1}(p_0) $, which is eliminated by $ \delta_{\pi_p^{-1}(p_0), p_0} \delta_{q, p_0} $. Now observe that every loop with unique patch index $ \alpha_j $ contains exactly one index $ p_j $ or $ p_j' $ on the left ($ \xi(j) = -1 $). So we get exactly one factor $ n_{\alpha_j, k}^2 $ for every $ \alpha_j $, except for $ n_{\alpha_q, k}^2 $. Thus,
\begin{equation}
	\begin{aligned}
	\mathrm{r.h.s.}
	= &\sum_{\substack{n = 2 \\ n : \mathrm{even}}}^\infty \frac{1}{2^n n!}
	\sum_{\xi} \sum_{(\pi_p, \pi_h)}
	\sum_k \frac{1}{n_{\alpha_q, k}^2} \sum_{\balpha \setminus \alpha_q}
	 \left( \prod_{j = 1}^n K(k)_{\alpha_j, \beta(\alpha_j)} \right) \sgn(\xi, \pi_p, \pi_h) \;.
\end{aligned}
\end{equation}
Note that the sum over $ k \in \Gamma^{\nor} $ here gets reduced to those $ k $ with $ \alpha_q \in \cI_k $ and $ q \mp k : \alpha_q $, as otherwise, the contribution vanishes. So comparing with \eqref{eq:cCq}, the sum becomes $ \sum_k = \sum_{k \in \tilde{\cC}^q \cap \ZZZ^3} $. As $ \beta $ is cyclic, the sum in $ \balpha \setminus \alpha_j $ over the $ K $-matrix elements amounts to an $ (n - 1) $-fold matrix multiplication, so
\begin{equation}
	\begin{aligned}
	\mathrm{r.h.s.}
	= &\sum_{\substack{n = 2 \\ n : \mathrm{even}}}^\infty \frac{1}{2^n n!}
	\sum_{\xi} \sum_{(\pi_p, \pi_h)}
	\sum_{k \in \tilde{\cC}^q \cap \ZZZ^3} \frac{1}{n_{\alpha_q, k}^2}
	 (K(k)^n)_{\alpha_q, \alpha_q} \sgn(\xi, \pi_p, \pi_h) \;.
\end{aligned}
\end{equation}
Next, we evaluate the sign factor $ \sgn(\xi, \pi_p, \pi_h) $, which is the same one needed to bring the entire diagram into maximally crossed form while ignoring $ a_q^* a_q $, see Fig.~\ref{fig:Friedrichs_multicommutator}. Observe that the maximally crossed form of the diagram obeys the bosonization structure, that is, it satisfies \eqref{eq:piconstraintstrict}. Every other bosonized diagram can be derived from it by a finite number of swaps of two $ c^\sharp $-vertices, each amounting to 4 swaps of fermionic connectors. So the total number of fermionic swaps is divisible by 4 and thus even, which immediately yields $ \sgn(\xi, \pi_p, \pi_h) = 1 $.\\
Finally, it remains to count the admissible diagrams, indexed by $ (\xi, \pi_p, \pi_h) $, contributing to the r.h.s.\;. For this, observe that the topological structure of all contributing diagrams is the same in the following sense: We can transform any diagram into any other by changing the order in which $ S_\pm $-vertices enter the diagram and swapping the two $ c^\sharp $-vertices inside certain $ S_\pm $-vertices. In total, there are $ 2^n $ ways to select in which of the $ n $ vertices $ S_\pm $ the two $ c^\sharp $-vertices shall be swapped. Next, consider the order in which the $ S_\pm $-vertices enter the diagram. At each contraction step $ j \in \{1, \ldots, n-1\} $, there are two $ S_\pm $-vertices which may join the existing diagram at that point, out of which one can be chosen. This renders $ 2^{n-1} $ distinct orders for the $ S_\pm $-vertices. Thus, $ \sum_\xi \sum_{(\pi_p, \pi_h)} $ yields $ 2^{2n-1} $ diagrams of identical value contributing to the r.h.s. of \eqref{eq:multicommutatorbos}. So comparing with \eqref{eq:nqb},
\begin{equation}
\begin{aligned}
	\mathrm{r.h.s.}
	= &\sum_{\substack{n = 2 \\ n : \mathrm{even}}}^\infty \frac{2^{n-1}}{n!}
	 	\sum_{k \in \tilde{\cC}^q \cap \ZZZ^3} \frac{1}{n_{\alpha_q, k}^2}
	 	(K(k)^n)_{\alpha_q, \alpha_q}\\
	 = &\frac{1}{2} \sum_{k \in \tilde{\cC}^q \cap \ZZZ^3} \frac{1}{n_{\alpha_q, k}^2} 
	 	\big( \cosh(2 K(k)) - 1 \big)_{\alpha_q, \alpha_q}
	 = n_q^{(\bos)} \;,
\end{aligned}
\end{equation}
which establishes \eqref{eq:multicommutatorbos}.\\

\end{proof}

\small
\noindent\textit{Acknowledgments.}
    This research was supported by the European Union (ERC FermiMath, grant agreement nr. 101040991 of Niels Benedikter). Views and opinions expressed are however those of the author(s) only and do not necessarily reflect those of the European Union or the European Research Council Executive Agency. Neither the European Union nor the granting authority can be held responsible for them. The author thanks Niels Benedikter for helpful discussions.
\normalsize

%
%

\end{document}